\documentclass[fleqn,usenatbib]{mnras}
\usepackage{newtxtext,newtxmath}
\usepackage[T1]{fontenc}
\DeclareRobustCommand{\VAN}[3]{#2}
\let\VANthebibliography\thebibliography
\def\thebibliography{\DeclareRobustCommand{\VAN}[3]{##3}\VANthebibliography}

\usepackage[dvipdfmx]{graphicx}	% Including figure files
\usepackage{amsmath}	% Advanced maths commands
\usepackage{color}
\usepackage{threeparttable}
\usepackage{array,booktabs}
\usepackage{siunitx}
\usepackage{subcaption}

\newcommand{\msun}{{\rm M_\odot}} %Msun
\newcommand{\zsun}{{\rm Z_\odot}} %Zsun
\newcommand{\yr}{{\rm yr}} %yr
\newcommand{\Myr}{{\rm Myr}} %Myr
\newcommand{\Gyr}{{\rm Gyr}} %Gyr
\newcommand{\pc}{{\rm pc}} %pc
\newcommand{\kpc}{{\rm kpc}} %kpc
\newcommand{\kelvin}{{\rm K}} %kelvin

\newcommand{\mstar}{M_\ast}
\newcommand{\sfr}{{\rm SFR}}
\newcommand{\oh}{{\rm 12+log(O/H)}}
\newcommand{\muv}{{\rm M_{UV}}}
\newcommand{\luv}{L_{\rm UV}}

\newcommand{\rmin}{R_{\rm min}}
\newcommand{\rmax}{R_{\rm max}}
\newcommand{\reff}{R_{\rm eff}}

\newcommand{\sgmg}{\Sigma_{\rm g}}
\newcommand{\sgms}{\Sigma_\ast}

\newcommand{\sgmsf}{\dot{\Sigma}_{\rm sf}}
\newcommand{\sgmin}{\dot{\Sigma}_{\rm in}}
\newcommand{\sgmout}{\dot{\Sigma}_{\rm out}}
\newcommand{\sgmret}{\dot{\Sigma}_{\rm ret}}
\newcommand{\sgmzr}{\dot{\Sigma}_{\rm Z,ret}}
\newcommand{\mret}{m_{\rm ret}}

\newcommand{\sgmdp}{\dot{\Sigma}_{\rm d}^{\rm pro}}
\newcommand{\sgmda}{\dot{\Sigma}_{\rm d}^{\rm ast}}
\newcommand{\sgmdd}{\dot{\Sigma}_{\rm d}^{\rm des}}
\newcommand{\sgmde}{\dot{\Sigma}_{\rm d}^{\rm eje}}
\newcommand{\sgmdg}{\dot{\Sigma}_{\rm d}^{\rm gro}}
\newcommand{\sgml}{I_{\rm UV}}

\newcommand{\fspin}{f_{\rm s}}
\newcommand{\fout}{\eta_{\rm 0}}
\newcommand{\mhch}{M_{\rm h,0}}
\newcommand{\fboost}{\mathcal{F}_{\rm b}}

\newcommand{\mmax}{m_{\rm max}}
\newcommand{\mmin}{m_{\rm min}}

\title[Galaxy evolution at $z > 5$]{Bridging Theory and Observations: Insights into Star Formation Efficiency and Dust Attenuation in $z > 5$ Galaxies}
\author[D.~Toyouchi et al.]{
Daisuke~Toyouchi$^{1}$\thanks{E-mail: d.toyouchi@gmail.com},
Hidenobu~Yajima$^{2}$,
Andrea~Ferrara$^{3}$,
Kentaro~Nagamine$^{1,4,5,6,7}$
\\
$^{1}$Theoretical Astrophysics, Department of Earth \& Space Science, Graduate School of Science, Osaka University, \\
1-1 Machikaneyama, Toyonaka, Osaka, 560-0043, Japan\\
$^{2}$Center for Computational Sciences, University of Tsukuba, 1-1-1 Tennodai, Tsukuba, Ibaraki 305-8577, Japan\\
$^{3}$Scuola Normale Superiore, Piazza dei Cavalieri 7, 50126 Pisa, Italy\\
$^{4}$Theoretical Joint Research, Forefront Research Center, Graduate School of Science, The University of Osaka, Toyonaka, Osaka 560-0043, Japan\\
$^{5}$Kavli IPMU (WPI), UTIAS, The University of Tokyo, Kashiwa, Chiba 277-8583, Japan\\
$^{6}$Department of Physics \& Astronomy, University of Nevada, Las Vegas, 4505 S. Maryland Pkwy, Las Vegas, NV 89154-4002, USA\\
$^{7}$Nevada Center for Astrophysics, University of Nevada, Las Vegas, 4505 S. Maryland Pkwy, Las Vegas, NV 89154-4002, USA
}

\date{Accepted XXX. Received YYY; in original form ZZZ}

\pubyear{2025}

\begin{document}
\label{firstpage}
\pagerange{\pageref{firstpage}--\pageref{lastpage}}
\maketitle

\begin{abstract}
We investigate early galaxy evolution by modeling self-consistently their radially-resolved evolution of gas, stars, heavy elements, and dust content. Our model successfully reproduces various observed properties of JWST-identified galaxies at $z > 5$, including sizes, stellar masses, star formation rates (SFR), metallicities, and dust-to-stellar mass ratios. 
We show that the star formation efficiency (SFE), $f_\ast \equiv \sfr/(f_{\rm b} \dot{M}_{\rm h})$, is regulated by the global equilibrium between cosmological gas inflows, star formation, and gas outflows. Our model predicts $f_\ast \lesssim 20~\%$ for galaxies with halo masses of $M_{\rm h} \sim 10^{11-12}\, \msun$ down to $z = 5$, allowing them to reach intrinsic UV magnitudes of $M_{\rm UV} \lesssim -22~{\rm mag}$; when dust attenuation is ignored, the predicted  UV luminosity function (LF) at $z \sim 12$ agrees well with observations. However, our model also suggests that these galaxies would be heavily obscured by dust, with high optical depths at 1500~\AA~of $\tau_{1500} \gtrsim 10$, 
causing the dust-attenuated UV LF to fall significantly below the observed one. This discrepancy highlights the need for mechanisms that mitigate strong dust attenuation, such as dust evacuation from star-forming regions and/or preferential production of large dust grains. Further exploration of these processes is essential for understanding the early stages of galaxy evolution.

\end{abstract}


%%%%%%%%%%%%%%%%%%%%%%%%%%%%%%%%%%%%%%%%%%%%%%%%%%

%%%%%%%%%%%%%%%%% BODY OF PAPER %%%%%%%%%%%%%%%%%%

\section{Introduction}\label{sec:intro}

The James Webb Space Telescope (\textit{JWST}) has significantly advanced our understanding of galaxy formation and evolution in the early universe. 
Its unprecedented capabilities provide a unique opportunity to investigate the properties of $z > 10$ galaxies that emerged within the first few hundred million years. 

A notable finding from JWST observations is a higher abundance of $z > 10$ galaxies than expected from theoretical models, particularly at the bright end of the UV luminosity functions (LFs) \citep[e.g.,][]{Yung2019MNRAS, Behroozi2020MNRAS, Wilkins2023MNRAS, Mason2023MNRAS}, which also exhibit a surprisingly gradual evolution across $z \sim 9$--$15$ 
\citep[e.g.,][]{Finkelstein2023ApJ, Peres-Gonzalez2023ApJ, McLeod2024MNRAS, Adams2024ApJ, Donnan2024MNRAS, Harikane2023ApJS, Harikane2024barXiv}.

The observed galaxy UV LFs provide key insights into redshift dependence on star formation efficiencies (SFEs).
Some theoretical studies have proposed a highly efficient star formation mode in super-early galaxies, driven by their extremely high gas densities leading to gas consumption timescales shorter than the lifetimes of massive stars \citep{Dekel2023MNRAS, Li2024A&A}. 
Indeed, \citet{Harikane2023ApJS} suggest that a constant SFE observed in $z \sim 2$-7 galaxies \citep{Harikane2022ApJS} is insufficient to explain cosmic star formation rates at $z > 10$, implying an increase in SFEs toward higher redshifts.
Conversely, \citet{Donnan2025arXiv} demonstrate with a simple model calculation that the observed UV LFs at $z > 10$ can be reproduced assuming an empirical stellar-to-halo mass relation derived at $z \simeq 0$, suggesting no significant evolution in SFEs across cosmic time.
This raises the key question: what mechanisms govern SFEs within galaxies, and whether an increase of SFEs with redshift is truly necessary.

An alternative explanation for the bright-end UV LFs at high redshift is a top-heavy initial mass function (IMF), which would enhance the UV emissivity per unit star formation rate (SFR) \citep[e.g.,][]{Inayoshi2022ApJ, Wang_Y2023ApJ, Trinca2024MNRAS, Hutter2024arXiv}.
For an IMF to be top-heavy, gas clouds in galaxies must have sufficiently low metallicities ($Z < 10^{-3}~Z_\odot$) to suppress gas fragmentation due to metal cooling \citep[e.g.,][]{Chon2024MNRAS}.
However, recent JWST observations suggest that UV-bright galaxies at $z > 10$ already exhibit metallicities of $\gtrsim 0.1~Z_\odot$
\citep[][]{Bunker2023AA, Carniani2024Natur, Carniani2024arXiv, D'Eugenio2024, Castellano2024ApJ}, 
implying that their IMFs might have already settled onto a standard Salpeter-like distribution.
Indeed, \citet{Cueto2024A&A} argue that variations in the IMF due to metallicity alone are unable to explain the observed UV LF.

If these super-early galaxies are so metal-enriched, they are also likely to be dust-rich. 
For instance, assuming that the dust-to-gas mass ratio scales with metallicity, the dust mass in these galaxies could have already reached $\sim 10^5~\msun$.
Moreover, given the compact sizes of $z > 10$ galaxies ($R_{\rm eff} \lesssim 200~\pc$), they could become highly optically thick to UV radiation due to dust attenuation, with optical depths at 1500~\AA~of $\tau_{1500} \gtrsim 10$ and V-band extinction of $A_V \gtrsim 4$. 
This expectation, however, contradicts extremely blue UV slopes of these galaxies exhibit extremely blue UV slopes, suggesting minimal attenuation \citep[$\beta < -2.2$, e.g.,][]{Topping2022ApJ, Topping2024MNRAS, Cullen2024MNRAS, Morales2024ApJ}.
To resolve this discrepancy, the Attenuation-Free Model (AFM) has been proposed, where radiation-driven outflows displace dust onto larger ($\sim $ kpc) scale, thereby reducing the dust optical depth by several orders of magnitude \citep[e.g.,][]{Ferrara2023MNRAS, Fiore2023ApJ, Ziparo2023MNRAS}.
Nevertheless, several aspects of the idea need deeper inspection, and it is therefore crucial to conduct a comprehensive analysis for such super-early galaxies that accounts for galaxy size and metallicity, dust mass, and their consistency with observed UV luminosities and number densities.

In this paper, we present a new galaxy evolution model that self-consistently incorporates galaxy size evolution, chemical enrichment, and dust formation. Using this model, we aim to reproduce various characteristics of JWST-identified galaxies at $z > 5$ and provide insights into key processes such as gas inflows and outflows that govern galaxy evolution.  Furthermore, we examine SFEs of galaxies as a function of host halo mass and redshift, and discuss the requirements for dust attenuation to maintain consistency with the observed UV LFs. Our findings provide crucial insights into the physical processes shaping the galaxy evolution and their observational signatures at $z > 5$.

This paper is structured as follows.
In \S~\ref{sec:model}, we describe the numerical setup of our galaxy evolution model.
In \S~\ref{sec:results}, we compare our model predictions with recent observations of $z > 5$ galaxies, focusing on galaxy size, stellar mass, SFRs, metallicity, and dust mass.
In \S~\ref{sec:SFE}, we present our main results on SFEs and their redshift dependence.
Based on these results, we discuss the UV LFs of high redshift galaxies and the necessity of reducing dust attenuation within galaxies in \S~\ref{sec:UV-LF}, and 
potential mechanisms for it are further explored in \S~\ref{sec:DA}.
Additionally, in \S~\ref{sec:fbf}, we examine a weak-feedback scenario that has been proposed to enhance SFEs for super-early galaxies.
Finally, we summarize our conclusion in \S~\ref{sec:summary}.
Throughout this paper, we use the Planck cosmological parameter sets: $\Omega_{\rm m} = 0.3111$, $\Omega_{\rm \Lambda} = 0.6899$, $\Omega_{\rm b} = 0.0489$, and $h = 0.6766$, and $\sigma_8 = 0.8102$ \citep[][]{Planck2020A&A}.

% We present a theoretical model of high-$z$ galaxies.
% Our model results generally explain observed properties of $z > 5$ galaxies, such as their sizes, stellar masses, star formation rates, metallicities, and dust contents.
% We also predict the UV luminosity function (LF) with our calculation results.
% Our analysis can be compared to \citet{Inayoshi2022ApJ},
% who estimated $\sfr$ and $\luv$ for each galaxy, assuming a star formation efficiency defined as $f_\ast \equiv \sfr / (f_{\rm b} \dot{M_{\rm h}})$.
% They concluded that $f_\ast \gtrsim 30~\%$, supposing a Salpeter-like IMF, is necessary to explain the bright end of the UV LF at $z > 13$.
% This star formation efficiency is far higher than the typical value for local star-forming galaxies, $f_\ast \simeq 0.01$ \citep[e.g.,][]{Bigiel2008AJ},
% and thus, its physical explanation remains an open question.
% It is worth noting here that the UV LF at $z \gtrsim 10$ was updated by \citet{Harikane2024barXiv, Harikane2024aApJ} based on spectroscopically-confirmed galaxies,
% showing that the number density at the bright end is reduced by a factor of a few compared to the UV LF used in \citet{Inayoshi2022ApJ}.
% This implies that the required star formation efficiency would be also lower.
% In this paper, we revisit this argument based on our detailed calculations and the latest UV LF observations.

%%%%%%%%%%%%%%%%%%%%%%%%%%%%%%%%%%%%%%%%%%%%%%%%%%

\section{Galaxy evolution model}\label{sec:model}

\subsection{Basic Equations}\label{sec:basic_eqs}

We calculate galaxy evolution from $z = 20$ to $z = 5$, corresponding to the cosmic age of $0.18~{\rm Gyr}$ and $1.2~{\rm Gyr}$, respectively.
We assume that galaxies already possess entirely rotation-supported disks \citep[$v/\sigma \gtrsim 1$, e.g.,][]{Parlanti2023A&A, Nelson2024ApJ, de_Graaff2024A&A, Xu2024ApJ, Kohandel2024A&A}
, conducting one-dimensional calculations that space the disk structure with a logarithmic radial grid covering a range from $\rmin = 1~\pc$ to $\rmax = 10^4~\pc$ with fifty cells.
At each radius, we calculate the time evolution of surface mass densities of gas, stars, and heavy elements, denoted as $\sgmg$, $\sgms$, and $Z \sgmg$, respectively.
% Here, we focus on oxygen as a representative of heavy elements.
The basic equations solved in this paper are written as follows:
\begin{eqnarray}
\frac{\partial \sgmg}{\partial t} = - \sgmsf + \sgmin - \sgmout + \sgmret \ ,
\label{eq:cons_g}
\end{eqnarray}
\begin{eqnarray}
\frac{\partial \sgms}{\partial t} = \sgmsf - \sgmret \ ,
\label{eq:cons_s}
\end{eqnarray}
\begin{eqnarray}
\frac{\partial (Z \sgmg)}{\partial t} = - Z \sgmsf + Z_{\rm in} \sgmin - Z \sgmout + \sgmzr \ ,
\label{eq:cons_z}
\end{eqnarray}
where $\sgmsf$, $\sgmin$, and $\sgmout$ are the surface densities of star formation rates, gas inflow rates, and gas outflow rates, respectively, whose specific expressions are introduced in the next subsection (\S~\ref{sec:SF_IF_OF}).
We set the metallicity of inflowing gas as $Z_{\rm in} = 0$, considering metal-free gas inflows for simplicity.
$\sgmret$ and $\sgmzr$ represent mass-return rates of gas and heavy elements via stellar evolution, which are described as follows:
\begin{eqnarray}
\sgmret = \int^{\mmax}_{m(t = t_{\rm l})} \mret(m, Z(t')) \phi(m, Z(t')) \sgmsf(t') {\rm d}m \ ,
\label{eq:sgmret}
\end{eqnarray}
\begin{eqnarray}
\sgmzr = \int^{\mmax}_{m(t = t_{\rm l})} \left ( m_{\rm II}(m, Z(t')) + Z(t') \mret(m, Z(t')) \right ) \nonumber \\ 
\times \phi(m, Z(t'))  \sgmsf(t') {\rm d}m \ ,
\label{eq:sgmzr}
\end{eqnarray}
where $m$ denotes the initial mass of progenitor stars in unit of $\msun$, 
$\phi$ is the IMF normalized as $\int^{\mmax}_{\mmin}m\phi(m){\rm d}m = 1$ ($\mmin = 0.04~\msun$ and $\mmax = 150~\msun$),
$\mret$ is the return mass from a star, 
$m_{\rm II}$ is the mass of heavy elements synthesized and released by a type-II supernova, 
$t_{\rm l}(m)$ is the main-sequence stellar lifetime,
and we define $t' \equiv t - t_{\rm l}(m)$.
We model the stellar lifetime with the fitting formula proposed by \citet{Schaller1992A&AS}:
\begin{eqnarray}
\tau(m) = \frac{2.5 \times 10^3 + 6.7 \times 10^2 m^{2.5} + m^{4.5}}{3.3 \times 10^{-2} m^{1.5} + 3.5 \times 10^{-1} m^{4.5} }~{\rm Myr} \ .
\label{eq:tau_m}
\end{eqnarray}
We evaluate $\mret$ and $m_{\rm II}$ based on the metallicity-dependent table provided by \citet{Nomoto2013ARA&A}.
Note here that we ignore the chemical enrichment from type Ia supernovae since it is expected to be inactive at $z > 5$ yet owing to the long delay time of $\sim 0.5~\Gyr$ \citep[e.g.,][]{Matteucci1986A&A, Matteucci2001ApJ, Maoz2010ApJ}.
Finally, we adopt a metallicity-dependent IMF derived from radiation hydrodynamics simulations by \citet{Chon2024MNRAS}, who investigated stellar cluster formation across varying gas metallicities.
The IMF is approximated as follows:
\begin{eqnarray}
\phi = \phi_0 m^{-\alpha} \left [ 1-{\rm exp} \left (- \left ( \frac{m}{m_0} \right )^c \right ) \right ] 
{\rm exp} \left ( -\frac{\mmin}{m} - \frac{m}{\mmax} \right ) \ ,
\label{eq:imf}
\end{eqnarray}
\begin{eqnarray}
\alpha = 2.3 + 0.33~{\rm log} \left ( \frac{Z}{Z_\odot} \right ) \ ,
\label{eq:alpha}
\end{eqnarray}
\begin{eqnarray}
{\rm log} \left ( m_0 \right ) = 0.2 + 0.45~{\rm log} \left ( \frac{Z}{Z_\odot} \right ) \ ,
\label{eq:m0}
\end{eqnarray}
where we fix $c = 1.6$ and $Z_\odot = 0.02$ is the metal-to-gas mass ratio of the Sun. 
This formulation predicts a top-heavy IMF with $\alpha \sim 1$ for $Z = 10^{-4}~Z_\odot$, which gradually turns into a Salpeter-like one with $\alpha = 2.3$ for $Z = Z_\odot$.

We acknowledge that  Eqs.~(\ref{eq:cons_g}) and (\ref{eq:cons_z}) do not account for radial gas advection across the galactic disk. This simplification is likely reasonable at very high redshifts, where the high gas inflow rates driven by the cosmological mass assembly predominantly determine the gas density at each radius rather than radial advection fluxes. However, as discussed in \S~\ref{sec:M-size}, some UV-bright galaxies identified at $z > 10$ are too compact to be explained without efficient angular momentum extraction, potentially driven by galactic disk instabilities. While our model demonstrates the formation of such compact galaxies by reducing the spatial extent of gas inflows phenomenologically (\S~\ref{sec:SF_IF_OF}), 
a self-consistent treatment of radial advection would be an important aspect to incorporate in future model calculations.

\subsection{Star formation, Inflow, and Outflow}\label{sec:SF_IF_OF}

Eqs.~(\ref{eq:cons_g}-\ref{eq:cons_z}) can be solved by specifying functional forms of $\sgmsf$, $\sgmin$, and $\sgmout$.
We model $\sgmsf$ based on the Kennicutt-Schmidt (KS) law \citep{Schmidt1959ApJ, Kennicutt1998ApJ}:
\begin{eqnarray}
\sgmsf = 2.5 \times 10^{-4}~\mathcal{F}_{\rm b}~\left ( \frac{\sgmg}{\msun~\pc^{-2}} \right )^{1.4}~\msun~{\rm Myr^{-1}}~\pc^{-2} \ ,
\label{eq:sgmsf}
\end{eqnarray}
where we introduce a boost factor $\mathcal{F}_{\rm b} \geq 1$ to represent the tension in $\sgmsf$ from the original KS law.
Recent observations have reported $\mathcal{F}_{\rm b} \sim 5$--$10$ for $z > 5$ galaxies \citep[e.g.,][]{Vallini2024MNRAS}.
This enhancement in $\sgmsf$ would result from the relatively high surface gas densities in high-$z$ galaxies, which allow star-forming gas clouds to survive from stellar feedback, such as UV photoionization by massive stars \citep[e.g.,][]{Kim2018ApJ, Fukushima2021MNRAS} and energy injection by supernovae \citep[e.g.,][]{Grudic2018MNRAS, Grudic2020MNRAS}.
However, it remains highly uncertain whether an elevated star formation law is essential to explain various observed properties of high-$z$ galaxies \citep[][]{Donnan2025arXiv}.
To account for this uncertainty, we explore a range of $\fboost = 1$--10 in our model, adopting a fiducial value of $\fboost = 1$ to provide a conservative assessment of SFEs.

Galactic gas inflows occur as a natural consequence of cosmological mass assembly.
In this study, we model the mass growth of dark matter halos with a fitting formula from \citet{Fakhouri2010MNRAS}, which represents the mean growth rate of halos with mass $M_{\rm h}$:
\begin{eqnarray}
\dot{M}_{\rm h} = 46.1~\msun~\yr^{-1}~\left( \frac{M_{\rm h}}{10^{12}~\msun} \right)^{1.1} \nonumber \\
\times (1+1.11~z) \sqrt{\Omega_{\rm m}(1+z)^3+\Omega_\Lambda} \ .
\label{eq:mdoth}
\end{eqnarray}
The total gas inflow rate is then expressed as,
\begin{eqnarray}
\dot{M}_{\rm in} = f_{\rm b} \epsilon_{\rm in} \dot{M}_{\rm h} \ ,
\label{eq:mdotin}
\end{eqnarray}
where the cosmic baryon fraction is $f_{\rm b} \equiv \Omega_{\rm b}/\Omega_{\rm m} \sim 0.157$.
The coefficient $\epsilon_{\rm in}$ accounts for phenomenological suppression of gas inflows driven by virial shock gas heating at the halo's outskirts \citep[e.g.,][]{Birnboim2003MNRAS, Keres2005MNRAS, Dekel2006MNRAS}. 
We model this suppression as,
\begin{eqnarray}
\epsilon_{\rm in} = (1+M_{\rm h}/M_{\rm h,ch})^{-1} \ {\rm with} \ M_{\rm h,ch} = 10^{12}~\msun \ .
\label{eq:eps_in}
\end{eqnarray}
This prescription effectively suppresses the formation of extremely massive and UV-bright galaxies (e.g., $M_\ast > 10^{12}~\msun$ and $M_{\rm UV} < -24~{\rm mag}$).

Then, motivated by the ubiquitous exponential stellar profiles observed in disk galaxies \citep[e.g.,][]{Freeman1970ApJ, MacArthur2003ApJ},
we model the internal distribution of gas inflows with an exponential radial profile, $\sgmin \propto {\rm exp}(-R/h_R)$, where the normalization is given by $\dot{M}_{\rm in} = 2 \pi \int \sgmin R {\rm d}R$.
The scale length is defined as,
\begin{eqnarray}
h_R \equiv \fspin \lambda_{\rm s} r_{\rm vir} \ ,
\label{eq:hR}
\end{eqnarray}
where $r_{\rm vir} \propto M_{\rm h}^{1/3}(1+z)^{-1}$ is the virial radius of the host halo, $\lambda_{\rm s}$ is the spin parameter of the dark-matter halo, and $\fspin$ represents the ratio of the angular momentum of inflowing gas to that of dark matter.

Cosmological N-body simulations by \citet{Bullock2001MNRAS} show that the spin parameter follows a log-normal distribution with a median value of $\overline{\lambda}_{\rm s} = 0.035$ and a standard deviation of $\sigma_{\lambda} = 0.5$. 
This implies that over $70\%$ of halos have spin parameters in the range of $0.02 \lesssim \lambda_{\rm s} \lesssim 0.06$.
For simplicity, we adopt a constant value of $\lambda_{\rm s} = \overline{\lambda}_{\rm s}$ throughout this paper.

On the other hand, the value of $\fspin$ can significantly deviate from unity due to baryonic processes.
Stellar feedback can deposit angular momentum into the inflowing gas \citep[e.g.,][]{Bekki2009ApJ, Gibson2013A&A}, 
while tidal torques from galaxy mergers can extract angular momentum \citep[e.g.,][]{Hernquist1989Natur, Hopkins2010MNRAS}.
To account for these effects, we explore a range $\fspin = 0.1$-2 in our model, adopting a fiducial value as $\fspin = 0.5$ to represent moderate angular momentum extraction.

Additionally, we assume that outflow rates from each radius scale with the local star formation rates, expressed as $\sgmout = \eta~\sgmsf$, whose the mass loading factor $\eta$ determines the outflow efficiency.
In this paper, we model $\eta$ with a double power law function, following the results of cosmological simulations by \citet{Muratov2015MNRAS} as,
\begin{eqnarray}
\eta = \fout~\left \{ 
\frac{1}{2} \left( \frac{M_{\rm h}}{\mhch} \right)^{-1.1} +~
\frac{1}{2} \left( \frac{M_{\rm h}}{\mhch} \right)^{-0.33} \right \}\ .
\label{eq:fout}
\end{eqnarray}
The change in the power-law indices generally corresponds to the transition from energy-conserving to momentum-conserving outflows.
Supposing that outflow velocities scale with the circular velocities as $V_{\rm w} \propto M_{\rm h}^{1/3}$,
the index of $-0.33$ for massive halos represents momentum-driven winds ($\eta \propto V_{\rm w}^{-1}$).
Conversely, the index of $-1.1$ for low-mass halos is similar to but slightly steeper than the expectation for energy-driven winds ($\eta \propto V_{\rm w}^{-2}$).
According to \citet{Muratov2015MNRAS}, the transition occurs at $M_{\rm h} \sim 10^{10}$--$10^{11}~\msun$, so that we adopt $\mhch = 10^{11}~\msun$ throughout this study.
On the other hand, we treat $\fout$, the normalization of $\eta$ at $\mhch$, as a free parameter. 
Previous hydrodynamic simulations have reported large uncertainties in $\fout$, with values ranging from unity to a few tens \citep[][]{Muratov2015MNRAS, Christensen2016ApJ, Christensen2018ApJ, Pandya2021MNRAS, Harada2023MNRAS}.
Therefore, we explore a broad range of $\fout = 1$-100 and adopt a fiducial value of $\fout = 5$, representing an intermediate case.

\subsection{Dust content}\label{sec:dust}

Our model incorporates the dust formation in galaxies by solving the following mass conservation equation,
\begin{eqnarray}
\frac{\partial (D \sgmg)}{\partial t} = \sgmdp - \sgmda - \sgmdd - \sgmde + \sgmdg \ ,
\label{eq:cons_d}
\end{eqnarray}
where the dust-to-gas mass ratio is $D \leq Z$.
The five terms on the right-hand side represent the following physical processes 
\footnote{Our model description of those dust physics is based on the semi-analytic model D{\scriptsize ELPHI} presented by \citet{Dayal2022MNRAS}, so we recommend readers refer to their paper for details.}:

\begin{enumerate}
\item {\it Dust production:} $\sgmdp = y_{\rm d} \gamma_{\rm II}$ is the dust production rate by SN II, where $y_{\rm d}$ is the dust yield per SN II, and $\gamma_{\rm II} \equiv \int^{m_{\rm II,max}}_{m_{\rm II,min}}\sgmsf(t')\phi(m, Z(t')){\rm d}m$ ($m_{\rm II,min} = 10~\msun$ and $m_{\rm II,max} = 40~\msun$) is the SN II occurrence rate per unit area and time.
Throughout this paper, we set $y_{\rm d} = 0.1~\msun$ while standard values are a few times higher \citep[e.g.,][]{Todini2001MNRAS, Bianchi2007MNRAS}.
We adopt this assumption because, compared with local galaxies, $z \gtrsim 5$ galaxies have denser ISM owing to their compactness, so newly created dust by SN II would be efficiently destroyed through intense SN reverse shocks \citep[e.g.,][]{Nozawa2006ApJ, Nozawa2007ApJ, Leniewska2019A&A, Slavin2020ApJ}.

\item {\it Dust astration:} $\sgmda = D \sgmsf$ is the dust astration rate, where we assume that gas and dust are perfectly mixed in star-forming regions.

\item {\it Dust destruction:} $\sgmdd = (1-X_{\rm c}) D \sgmg / \tau_{\rm des}$ is the destruction rate of preexisting dust by SN shock, where $X_{\rm c}$ is the mass fraction of cold ISM, which can avoid dust destruction, and we assume $X_{\rm c} = 0.5$, based on recent high-$z$ galaxy simulations by \citet{Pallottini2019MNRAS}.
The dust destruction timescale is expressed as,
\begin{eqnarray}
\tau_{\rm des} =  \frac{\sgmg}{\epsilon \gamma_{\rm II} m_{\rm swep}} \ ,
\label{eq:tau_des}
\end{eqnarray}
where we adopt the dust destruction efficiency of $\epsilon = 0.03$ and 
the ISM mass swept by SN shock of $m_{\rm swep} = 6.8 \times 10^3~\msun$ \citep[][]{McKee1989ApJ, Lisenfeld1998ApJ}.

\item {\it Dust ejection:} $\sgmde = D \sgmout$ is the dust ejection rate via galactic outflows,
where we assume that outflows have the same dust-to-gas mass ratio as that in ISM since dust likely couples with gas strongly by Coulomb and viscous drag forces.

\item {\it Dust growth:} $\sgmdg$ is the dust growth rate via accretion of heavy elements in cold ISM,
which is written as
\begin{eqnarray}
\sgmdg =  \left ( 1- \frac{D}{Z} \right ) \frac{X_{\rm c} D \sgmg}{\tau_{\rm acc}} \ .
\label{eq:sgmdg}
\end{eqnarray}
The accretion timescale inversely scales with gas metallicity as $\tau_{\rm acc} = \tau_{0} (Z/Z_\odot)^{-1}$,
where the normalization timescale $\tau_0$ is usually set to be 1--100 Myr.
In this study, we adopt  $\tau_0 = 5~\Myr$ as the fiducial value to explain the dust-deficiency of $z > 10$ galaxies and the dust-richness of $z \sim 7$ galaxies simultaneously, as shown in \S~\ref{sec:Mdust}.

\end{enumerate}

\subsection{Stellar UV radiation and FIR dust continuum}\label{sec:Luv}

Based on our calculation results, we estimate the UV surface brightness $\sgml$ at any radius as,
\begin{eqnarray}
\sgml = \frac{1}{2} \epsilon_{\rm UV}~\sgmsf~c^2 \ ,
\label{eq:sgml}
\end{eqnarray}
where a factor of 1/2 accounts for radiation emitted from both surfaces of the galactic disk,
and $\epsilon_{\rm UV}$ is the UV radiative efficiency, which increases with top-heavier IMFs.
To model $\epsilon_{\rm UV}$, we fit the metallicity-dependent IMFs of \citet{Chon2024MNRAS}, obtaining the following relation:
\begin{eqnarray}
{\rm log} \left ( \epsilon_{\rm UV} \right ) = {\rm log} \left ( \epsilon_{\rm UV,0} \right ) + A \left \{ 1 + {\rm exp} \left ( \frac{{\rm [O/H]-[O/H]_0}}{\sigma_{\rm [O/H]}} \right ) \right \}^{-1} \ ,
\label{eq:eps_uv}
\end{eqnarray}
where ${\rm log} \left ( \epsilon_{\rm UV,0} \right ) = -3.463$, ${\rm [O/H]_0} = -1.308$, $\sigma_{\rm [O/H]} = 0.335$, and $A = 0.601$.

This formulation predicts a sharp decline in $\epsilon_{\rm UV}$ above ${\rm [O/H]} \sim -2$, leading to a transition in UV luminosities.
Specifically, at a fixed SFR, galaxies with ${\rm [O/H]} < -2$ appear about three times brighter in the UV band compared to those with ${\rm [O/H]} > -2$.
However, we note that the top-heavy IMF in such metal-poor galaxies enhances supernova feedback, which in turn reduces SFRs.
Semi-analytic calculations by \citet{Cueto2024A&A} demonstrated that, due to this counteracting effect, metallicity-dependent IMFs alone cannot fully explain the observed abundance of UV-bright galaxies at $z > 10$, a finding consistent with the recent cosmological galaxy formation simulations by \citet{Oku2024ApJ}.
While our model does not take into account the metallicity dependence of stellar feedback, such as variations in the mass loading factor ($\eta$), self-consistent modeling of both $\epsilon_{\rm UV}$ and $\eta$ as functions of metallicity remains an important subject for future research.

A fraction of UV photons are absorbed by dust grains within the galactic disk, so that the observed UV luminosity is reduced from the intrinsic value.
We account for the dust attenuation by calculating dust optical depth at 1500~\AA~ at each radius as,
\begin{eqnarray}
\tau_{\rm 1500} = \kappa_{1500} D \sgmg \sim 17.3~\left ( \frac{\sgmg}{10^{3}~\msun~\pc^{-2}} \right ) \left ( \frac{D/D_{\rm MW}}{0.1} \right ) \ , 
\label{eq:tau_uv}
\end{eqnarray}
where $D_{\rm MW} = 6.2 \times 10^{-3}$ is the dust-to-gas mass ratio in the Milky Way (MW) galaxy, and $\kappa_{\rm 1500} = 1.26 \times 10^{5}~{\rm cm^2 g^{-1}}$ is the dust mass absorption coefficient appropriate for a MW-like extinction curve \citep{Weingartner2001ApJ}.
Assuming a slab-like geometry, the escape fraction of UV continuum at 1500~\AA \,is written by
\begin{eqnarray}
f_{\rm esc} = \frac{1-{\rm exp}(-\tau_{\rm 1500})}{\tau_{\rm 1500}} \ ,
\label{eq:f_esc}
\end{eqnarray}
and the attenuated UV luminosity is derived by integrating the surface brightness over the disk,
\begin{eqnarray}
\luv = 4 \pi \int^{\rmax}_{\rmin} f_{\rm esc} \sgml R {\rm d}R \ .
\label{eq:luv}
\end{eqnarray}
This equation also describes the intrinsic UV luminosity for $f_{\rm esc} = 1$.

Dust grains that absorb UV photons emit the gained energy as continuum radiation in the FIR band.
The FIR flux of dust continuum reflects the intrinsic UV luminosity and the spatial distribution of dust in the galaxy.
Here, we compute the FIR flux at a rest frame wavelength of 158~$\micron$ since high-$z$ ALMA observations are often tuned to detect [CII] 158~$\micron$ line emission.
By considering the energy conservation, we model the local dust temperature at any radius as follows \citep[e.g.,][]{Dayal2010MNRAS, Hirashita2014MNRAS, Ferrara2022MNRAS}:
\begin{eqnarray}
T_{\rm d} = \left ( \frac{I_{\rm abs}}{\Theta D \sgmg} \right )^{1/(4+\beta_{\rm d})} \ ,
\label{eq:Td}
\end{eqnarray}
where 
\begin{eqnarray}
\Theta = \frac{8\pi}{c^2} \frac{\kappa_{158}}{\nu_{158}^{\beta_{\rm d}}}
\frac{k_{\rm B}^{4+\beta_{\rm d}}}{h_{\rm P}^{3+\beta_{\rm d}}} \Gamma (4+\beta_{\rm d}) \ ,
\label{eq:Theta}
\end{eqnarray}
and $I_{\rm abs} = (1-f_{\rm esc})\sgml$ is the stellar UV radiation absorbed by dust.
Here, we suppose the MW-like dust model, whose mass absorption coefficient is approximated as $\kappa_\nu = \kappa_{158}(\nu/\nu_{158})^{\beta_{\rm d}}$ with $\beta_{\rm d} = 2.03$,
$\kappa_{158} = 10.41~{\rm cm^{2} g^{-1}}$, and $\nu_{158} = c/({\rm 158~\mu m})$.
$\zeta$ and $\Gamma$ are the Zeta and Gamma functions, respectively,
and other symbols have the usual meaning.
Additionally, we account for dust heating by the CMB radiation, whose temperature is $T_{\rm CMB}(z)=T_0(1+z)$ with $T_0 = 2.7255~\kelvin$, by correcting the dust temperature with the following formula presented by \citet{Da_Cunha2013ApJ},
\begin{eqnarray}
T^{\prime}_{\rm d} = \left \{ 
T_{\rm d}^{4+\beta_{\rm d}}+T_0^{4+\beta_{\rm d}} [(1+z)^{4+\beta_{\rm d}}-1]
\right \}^{1/({4+\beta_{\rm d}})} \ .
\label{eq:Td2}
\end{eqnarray}

Finally, the total flux at the rest frame 158 $\micron$ from a galaxy is evaluated as,
\begin{eqnarray}
F_{158} = 4 \pi \int^{\rmax}_{\rmin} I_{158} R {\rm d} R \ ,
\label{eq:FIR}
\end{eqnarray}
where
\begin{eqnarray}
I_{158} = \frac{1}{2} g(z) \kappa_{158} D \sgmg [B_{158}(T^{\prime}_{\rm d}) - B_{158}(T_{\rm CMB}) ] \ ,
\label{eq:I_IR}
\end{eqnarray}
and $B_{\lambda}$ is the black-body spectrum, and $g(z) = (1+z)/d_{\rm L}^2$ with the luminosity distance to the source $d_{\rm L}$.

Thus, our model calculates the UV luminosity and the FIR flux self-consistently.
It is worth noting here that the FIR dust continuum radiation from $z > 10$ UV-bright galaxies has not been detected yet \citep[e.g.,][]{Fudamoto2024MNRAS, Carniani2024arXiv, Zavala2024arXiv, Schouws2024arXiv}, 
even though their observed half-light radii are so small that the optical depth can be extremely high, $\tau \gtrsim 10$ \citep[e.g.,][]{Ziparo2023MNRAS, Ferrara2024arXiv}.
This observational fact provides an important indication of the dust content and its spatial distribution in such super-early galaxies.
In \S~\ref{sec:Mdust} and \ref{sec:UV-LF}, we address this point in more detail based on our model calculation.

%%%%%%%%%%%%%%%%%%%%%%%%%%%%%%%%%%%%%%%%%%%%%%%%%%

%%%%%%%%%%%%% table:para %%%%%%%%%%%%%%
\renewcommand{\arraystretch}{1.5}
\begin{table*}
\begin{center}
\begin{tabular}[c]{ccclc} \hline \hline
Parameter & Range  & Fiducial value & Description \\ \hline
$\fboost$ & 1--10  & 1        & Boost factor that representing the enhancement in $\sgmsf$ from the original KS law, defined by Eq~(\ref{eq:sgmsf}) \\
$\fspin$  & 0.1--2 & 0.5      & Ratio of angular momentum of inflowing gas to dark matter, defined by Eq.~(\ref{eq:hR}) \\
$\fout$   & 1--100 & 5        & Outflow mass loading factor for $M_{\rm h,0} = 10^{11}~\msun$, defined by Eq.~(\ref{eq:fout}) \\
$\tau_0$  & 5--50~Myr & 5~Myr & Timescale of metal accretion onto dust at $Z = Z_\odot$, defined by Eq.~(\ref{eq:sgmdg}) \\
% $M_{\rm h}(z=5)$  & $10^{10}$--$10^{14}~\msun$ & $10^{12}~\msun$ & the dark matter halo mass at $z = 5$, with mass growth described by Eq.~(\ref{eq:mdoth}) \\
\hline \hline \\
\end{tabular}
\end{center}
\caption{
The four free parameters in our model. We present, from left to right, their symbols, explored ranges, fiducial values, and descriptions.
}
\label{table:para}
\end{table*}
\renewcommand{\arraystretch}{1}
%%%%%%%%%%%%%%%%%%%%%%%%%%%%%%%%%%%%%

%%%%%%%%%% fig:Mh_z %%%%%%%%%%%%%%%%
\begin{figure}
\centering
\includegraphics[width=0.9\columnwidth]{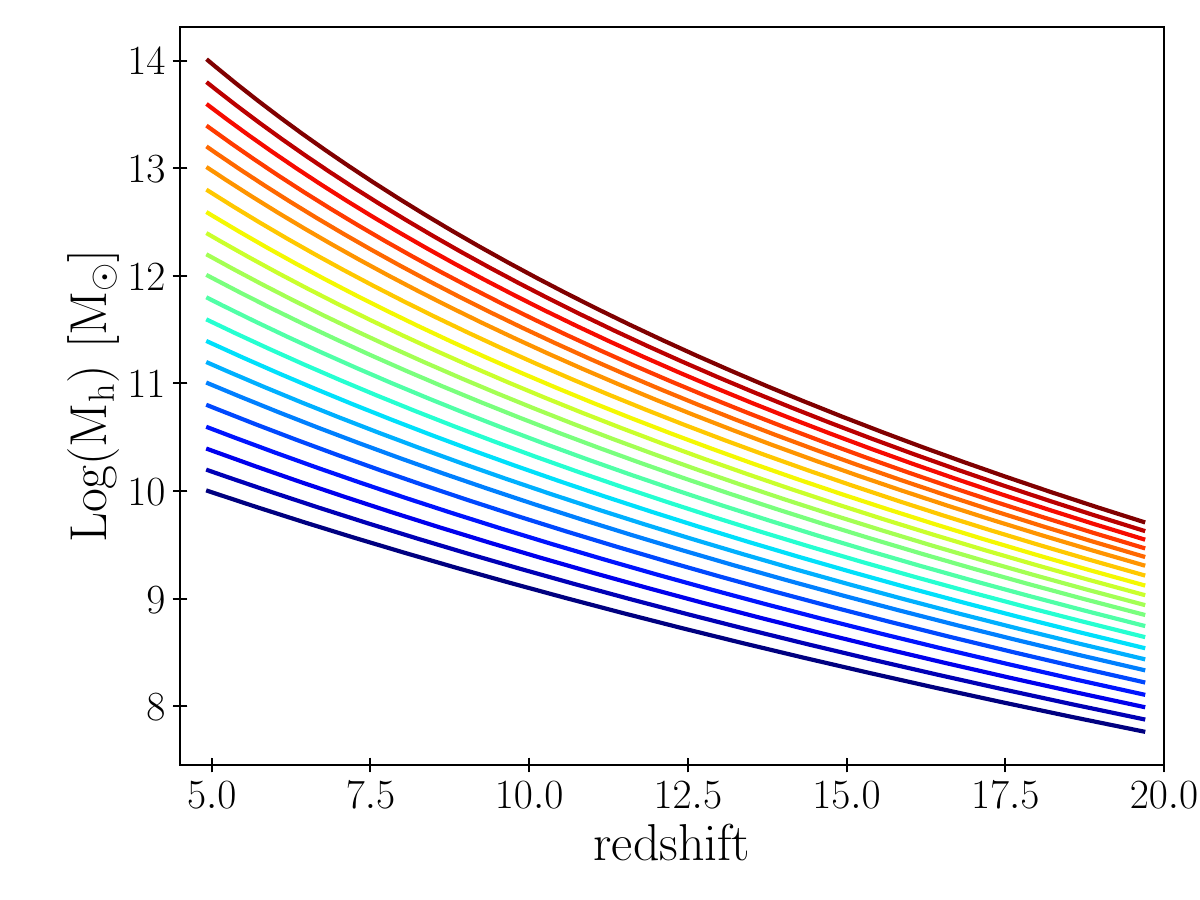}
\caption{
The redshift evolution of $M_{\rm h}$ for our 21 dark-matter halos.
The initial masses of the halos are spaced on a uniform logarithmic grid, ranging from $M_{\rm h} = 5.8 \times 10^{7}~\msun$ and $5.2 \times 10^{9}~\msun$ at $z = 20$.
The halo masses evolve with Eq.~(\ref{eq:mdoth}), and the lightest and heaviest halos grow to be $M_{\rm h} = 10^{10}~\msun$ and $10^{14}~\msun$ at $z = 5$, respectively.
}
\label{fig:Mh_z}
\end{figure}
%%%%%%%%%%%%%%%%%%%%%%%%%%%%%%%%%%

%%%%%%%%%% fig:mass %%%%%%%%%%%%%%%%
\begin{figure}
\centering
\includegraphics[width=0.9\columnwidth]{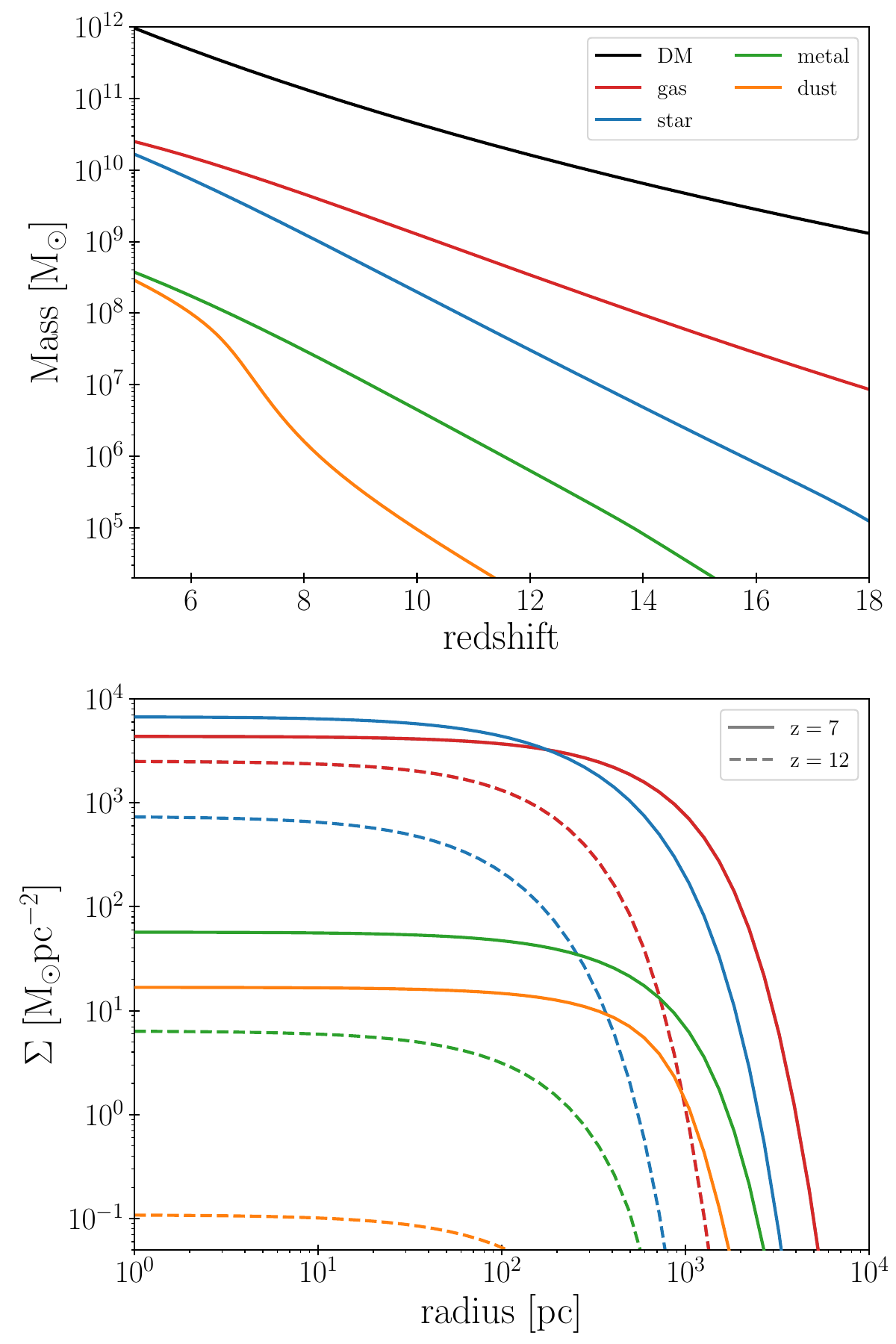}
\caption{
The mass evolution of different components for a single galaxy with $M_{\rm h}(z = 5) = 10^{12}~\msun$, calculated with the fiducial parameters: $\fboost = 1$, $\fspin = 0.5$, $\fout = 5$, and $\tau_{0} = 5~\Myr$.
The upper panel shows total masses of each component within the galaxy as a function of redshifts.
The masses of dark matter, gas, star, metal, and dust are represented by black, red, blue, green, and orange curves, respectively.
The lower panel shows the radial profiles of surface mass densities for each component in the galaxy.
Solid and dashed curves correspond to the profiles at $z = 7$ and $12$, respectively.
The line colors are consistent with those used in the upper panel.
}
\label{fig:mass}
\end{figure}
%%%%%%%%%%%%%%%%%%%%%%%%%%%%%%%%%%

\section{Results}\label{sec:results}

\subsection{Overview of our calculations}\label{sec:overview}

We apply our model to 21 dark matter halos, whose redshift evolution of $M_{\rm h}$ is calculated using Eq.~(\ref{eq:mdoth}) and summarized in Figure~\ref{fig:Mh_z}.
From the bottom to the top, we set the initial mass of $i$-th halo as $M_{\rm h, i} = M_{\rm h,1} \times 10^{\delta_i}$ with $\delta_{i} = (i-1)/20 \times {\rm log}(M_{\rm h,21}/M_{\rm h,1})$, where 
$M_{\rm h,1} = 5.8 \times 10^{7}~\msun$ and $M_{\rm h,21} = 5.2 \times 10^{9}~\msun$ correspond to the initial masses of the 1st (lightest) and 21st (heaviest) halos.
This setup ensures that the 1st, 5th, 9th, 13th, 17th, and 21st halos grow up to $M_{\rm h} = 10^{10},~10^{11},~10^{12},~10^{13}$ and $10^{14}~\msun$ at $z = 5$, respectively.
By incorporating these diverse halo mass growth histories, we can examine the galaxy evolution as a function of redshift and their host halo masses.

In this work, we perform model calculations by varying four parameters, $\fboost$, $\fspin$, $\fout$, and $\tau_0$. 
The ranges and fiducial values adopted in this study are summarized in Table~\ref{table:para}.
Comparing our model results with observations under different parameter sets provides valuable insights into star formation, gas inflows, gas outflows, and dust growth.

The upper panel of Figure~\ref{fig:mass} shows the mass evolution of different components for a halo with $M_{\rm h}(z = 5) = 10^{12}~\msun$, calculated with the fiducial parameters.
The gas mass increases over time, scaling approximately with the halo mass growth, with a gas-to-dark matter mass ratio of $M_{\rm g}/M_{\rm h} \sim 0.01$.
This value is about ten times lower than the cosmological baryon fraction, reflecting high outflow rates of $\dot{M}_{\rm out}/\sfr > 10$.
The stellar mass is initially much lower than but gradually catches up with the gas mass, resulting in a stellar mass fraction of $M_\ast/(M_{\rm g}+M_\ast) \sim 0.3$ at $z = 5$.
The mass evolution of heavy elements closely follows that of stellar mass, reflecting metal enrichment primarily driven by Type-II SNe.
A notable feature is the rapid dust mass growth: while significantly lower than the metal mass at $z > 10$, the dust mass reaches approximately $70~\%$ of the total metal mass by $z \sim 5$.
This sharp increase in dust mass for $z < 8$ is primarily driven by metal accretion on dust grains, as discussed in more detail in \S~\ref{sec:dust}.

The lower panel of Figure~\ref{fig:mass} presents the radial profiles of surface mass densities for the fiducial galaxy at $z = 7$ and $12$.
At $z = 12$, the gas radial profile exhibits an exponential decay beyond $R \sim 200~\pc$, marking the outer edge of the galactic disk; this cutoff radius increases to  $R \sim 1~\kpc$ by $z = 7$.
Despite the size growth, the gas mass density within the disk remains nearly constant at $\sgmg \sim 2 \times 10^3~\msun~\pc^{-2}$.
In contrast, the stellar mass density gradually increases, eventually surpassing the gas mass density by $z \sim 7$.

Simultaneously, the surface mass densities of metals and dust reach $Z \sgmg \sim 20~\msun~\pc^{-2}$ and $D \sgmg \sim 2~\msun~\pc^{-2}$, respectively.
The resulting dust-to-gas mass ratio of $D \sim 0.2~D_{\rm MW}$ corresponds to $\tau_{1500} \sim 70$, as given by Eq.~(\ref{eq:tau_uv}).
This implies that the galaxy is heavily obscured in the UV by dust.
In the following subsections, we further examine our results and discuss various observed properties of $z > 5$ galaxies.

%%%%%%%%%% fig:Ms_Mh %%%%%%%%%%%%%%%%
% \begin{figure}
% \centering
% \includegraphics[width=0.9\columnwidth]{figure/Ms_Mh.pdf}
% \caption{
% The stellar-to-halo mass ratio obtained by our fiducial model is shown with the red curve.
% }
% \label{fig:Ms_Mh}
% \end{figure}
%%%%%%%%%%%%%%%%%%%%%%%%%%%%%%%%%%

%%%%%%%%%% fig:Reff %%%%%%%%%%%%%%%%
\begin{figure}
    \centering
    \begin{subfigure}{\columnwidth}
        \centering
        \includegraphics[width=0.9\columnwidth]{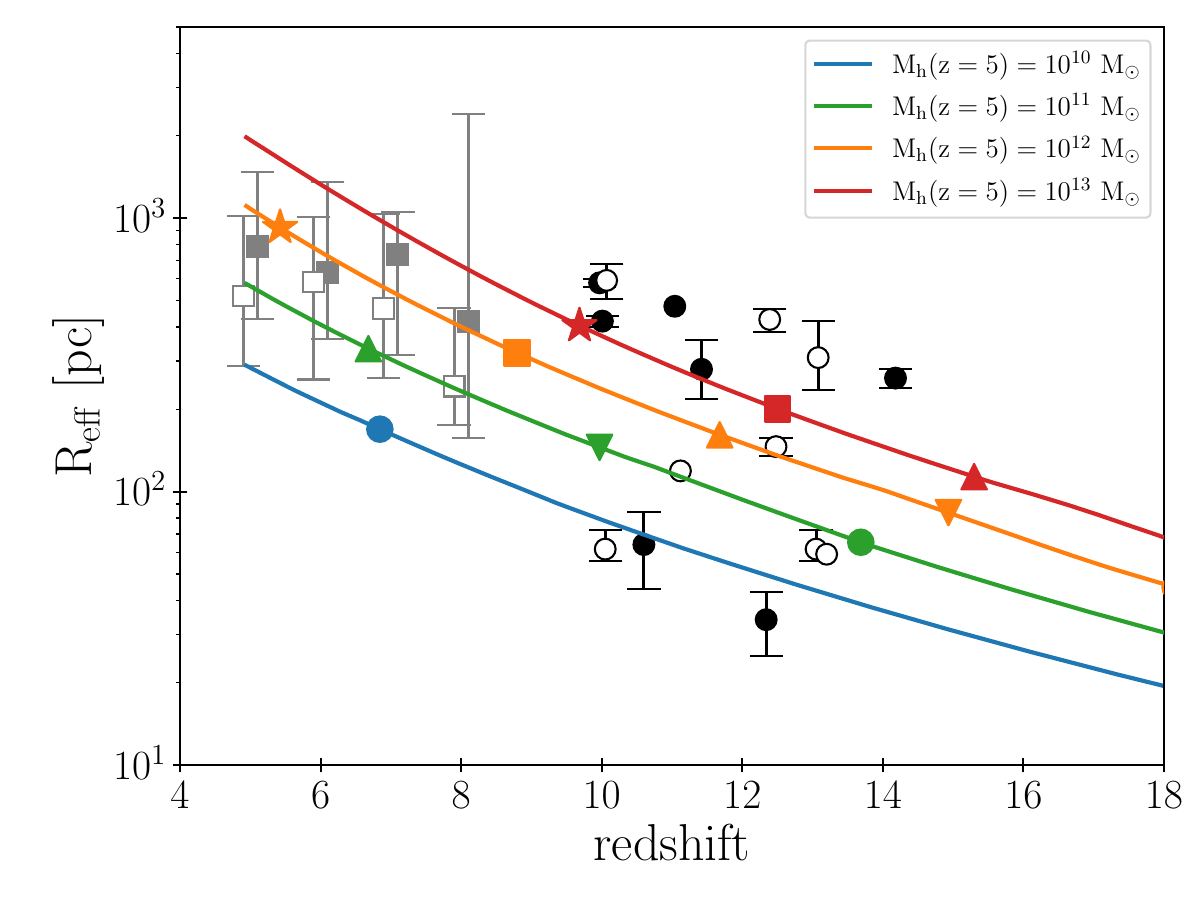}
    \end{subfigure}
    \begin{subfigure}{\columnwidth}
        \centering
        \includegraphics[width=0.9\columnwidth]{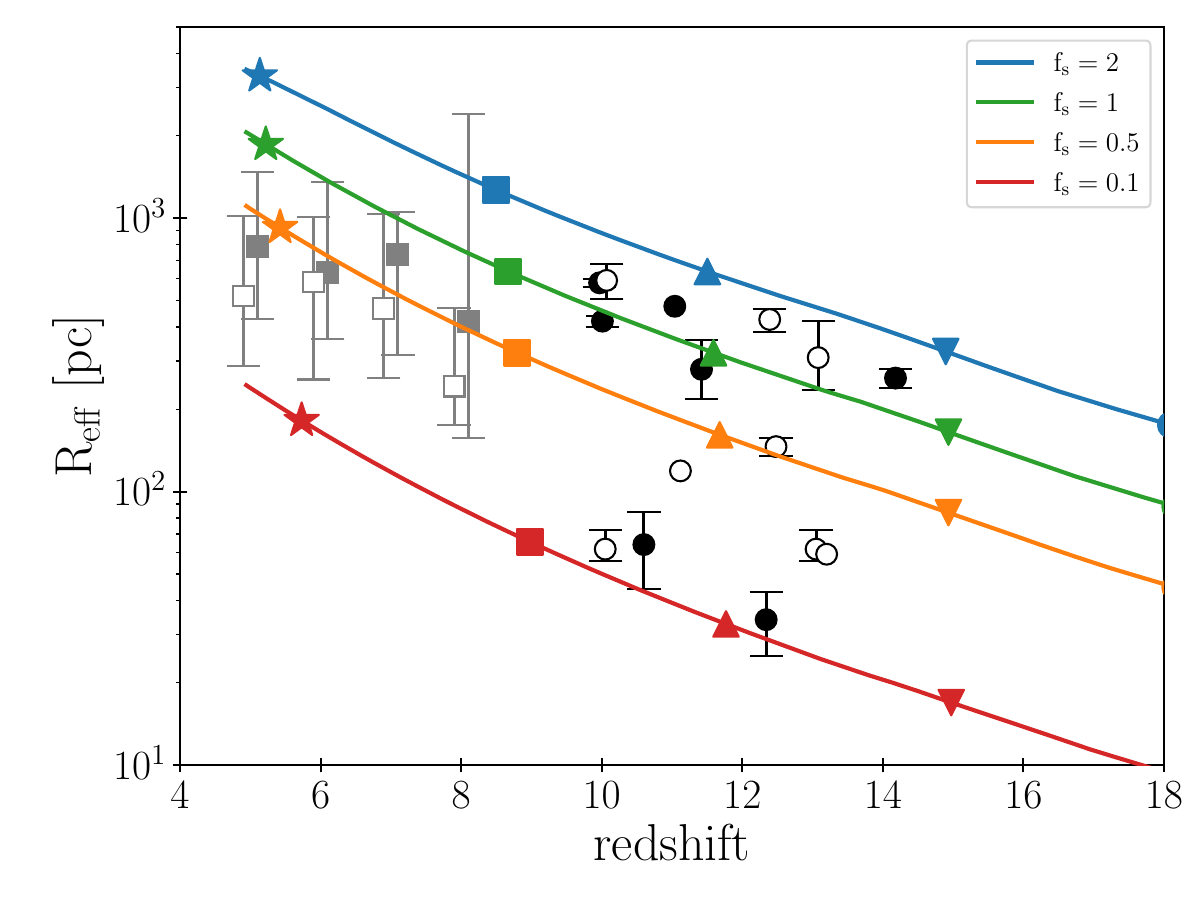}
    \end{subfigure}
\caption{
The redshift evolution of the UV half-light radii of galaxies predicted by our calculations.
In the upper panel, the blue, green, orange, and red curves show predictions from our fiducial model for four halos evolving to $M_{\rm h}(z = 5) = 10^{10},~10^{11},~10^{12},$ and $10^{13}~\msun$, respectively.
The lower panel presents results for a halo with $M_{\rm h}(z = 5) = 10^{12}~\msun$, calculated with different gas angular momentum parameters of $\fspin = 0.1$ (red), 0.5 (orange), 1 (green), and 2 (blue).
All other parameters not specified here are set to their fiducial values.
Colored markers indicate the redshifts where the intrinsic UV absolute magnitude reaches $\muv = -15,~-17,~-19,~-21$, and $-23$ mag, represented by circles, inverted triangles, triangles, squares, and stars, respectively.
For comparison, we also plot the half-light radii of Lyman-break galaxies with $\muv \sim -19~{\rm mag}$ and $-21~{\rm mag}$ at $5 < z < 8$ with gray open and filled squares, respectively, where values are averaged over redshift bins \citep[][]{Shibuya2015ApJS}.
Additionally, the black open and filled circles correspond to galaxies with $\muv > -20~{\rm mag}$ and $< -20~{\rm mag}$, respectively, identified by JWST at $z \gtrsim 10$ (see Table~\ref{table:sample} for references).
}
\label{fig:Reff}
\end{figure}
%%%%%%%%%%%%%%%%%%%%%%%%%%%%%%%%%%

%%%%%%%%%% fig:ms_sfr_z %%%%%%%%%%%%%%%%
\begin{figure*}
    \centering
    \begin{subfigure}{\textwidth}
        \centering
        \includegraphics[width=\textwidth]{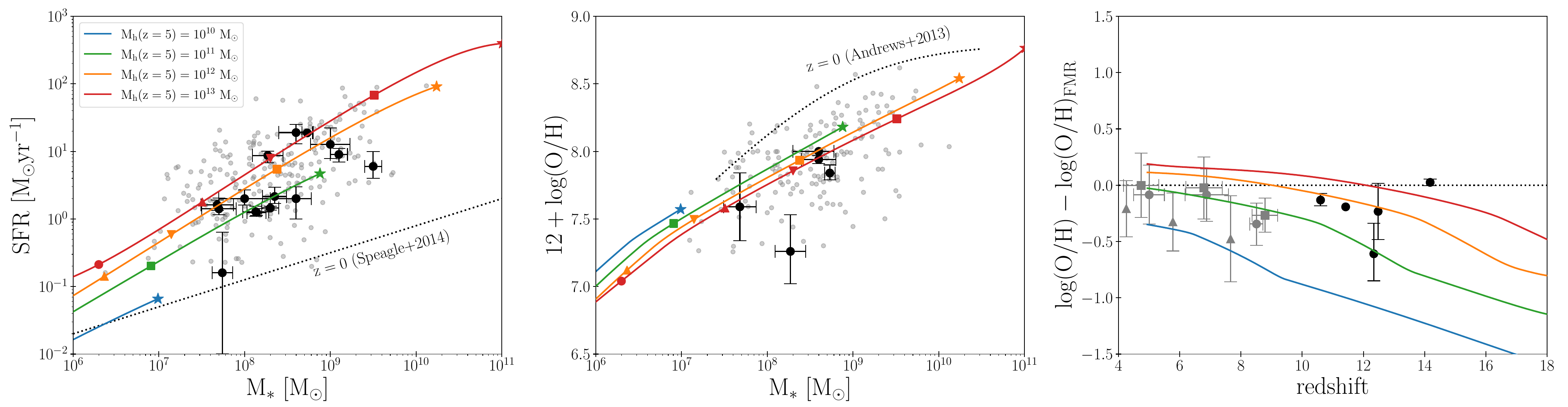}
    \end{subfigure}
    \begin{subfigure}{\textwidth}
        \centering
        \includegraphics[width=\textwidth]{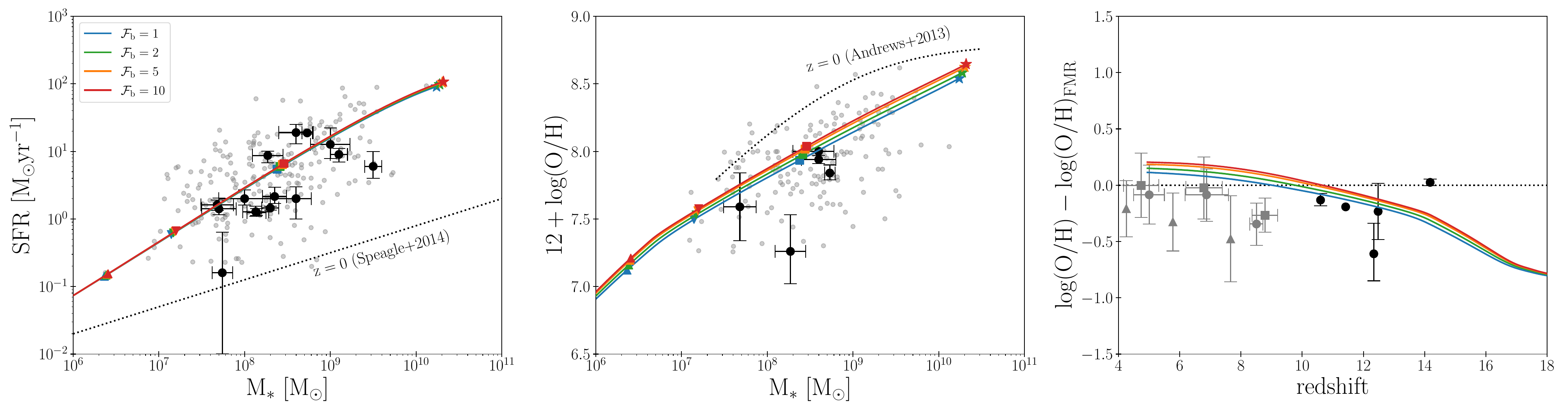}
    \end{subfigure}
    \begin{subfigure}{\textwidth}
        \centering
        \includegraphics[width=\textwidth]{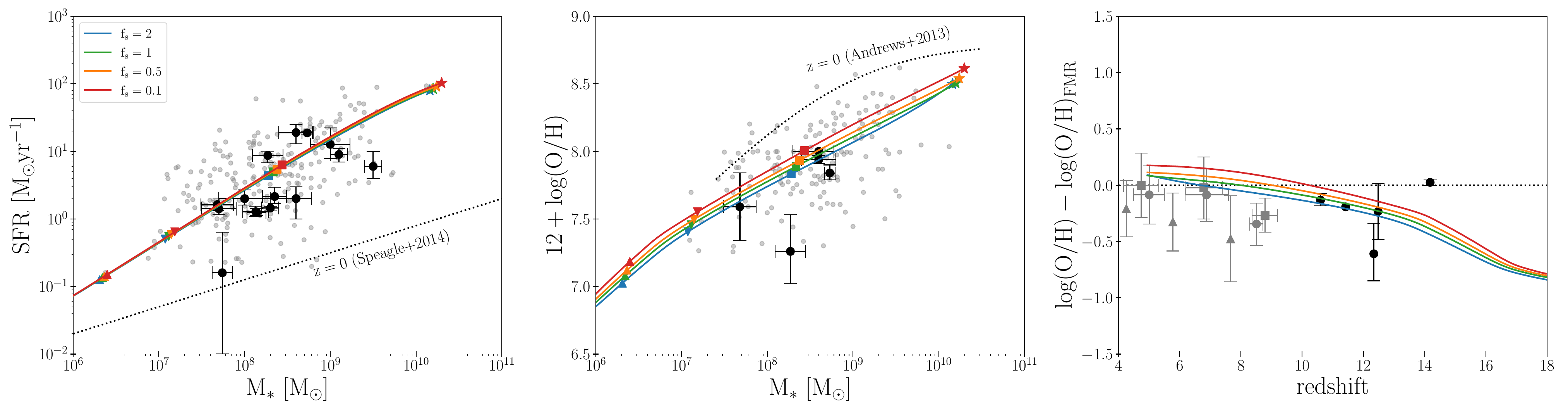}
    \end{subfigure}
    \begin{subfigure}{\textwidth}
        \centering
        \includegraphics[width=\textwidth]{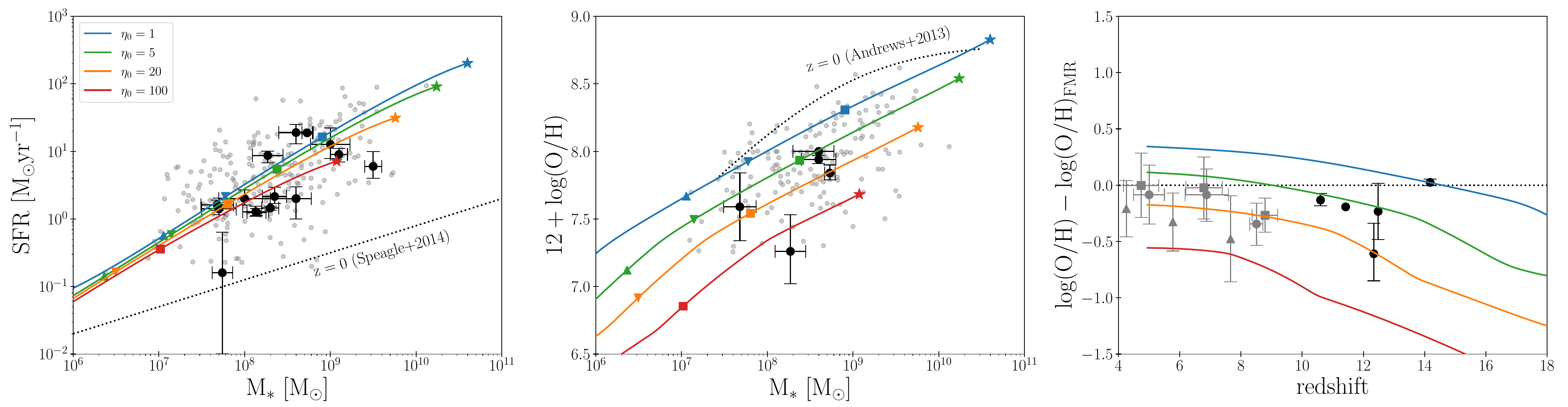}
    \end{subfigure}
\caption{
The relations between stellar mass, star formation rate, and metallicity obtained from our calculations.
{\it Left column}: The $\mstar$--$\sfr$ relation. 
{\it Middle column}: The $\mstar$--$\oh$ relations. 
{\it Right column}: The deviation in $\oh$ from the fundamental metallicity relation derived by \citet{Andrews2013ApJ} at any redshift.
In the top row, the blue, green, orange, and red curves show predictions from our fiducial model for four halos evolving to $M_{\rm h}(z = 5) = 10^{10},~10^{11},~10^{12},$ and $10^{13}~\msun$, respectively.
From the second to the fourth rows, the colored curves in each panel represent results for a halo with $M_{\rm h}(z = 5) = 10^{12}~\msun$, calculated with varying $\fboost$, $\fspin$, and $\fout$. 
Specific values for each parameter are noted in the top-left corner of the left panels,
while other parameters not specified there are set to their fiducial values.
In the left and middle column, colored markers denote redshifts $z = 18,~15,~13,~10$, and $5$, represented by circles, triangles, inverted triangles, squares, and stars, respectively.
For comparison, we plot observational data for star-forming galaxies at $3 < z < 10$ \citep[gray dots;][]{Nakajima2023ApJS, Curti2024A&A, Sarkar2024arXiv} and galaxies at $z > 10$ (black circles; see Table~\ref{table:sample} for references).
In the right column, gray circles, triangles, and squares represent redshift-binned values for $3 < z < 10$ galaxies from \citet{Nakajima2023ApJS}, \citet{Curti2024A&A}, and \citet{Sarkar2024arXiv}, respectively.
Additionally, the $\mstar$--$\sfr$ and $\mstar$--$\oh$ relations observed at $z = 0$ are shown as the dotted curves \citep{Speagle2014ApJS, Andrews2013ApJ}.
}
\label{fig:ms_sfr_z}
\end{figure*}
%%%%%%%%%%%%%%%%%%%%%%%%%%%%%%%%%%

%%%%%%%%%% fig:Md_z %%%%%%%%%%%%%%%%
\begin{figure}
\centering
\includegraphics[width=0.9\columnwidth]{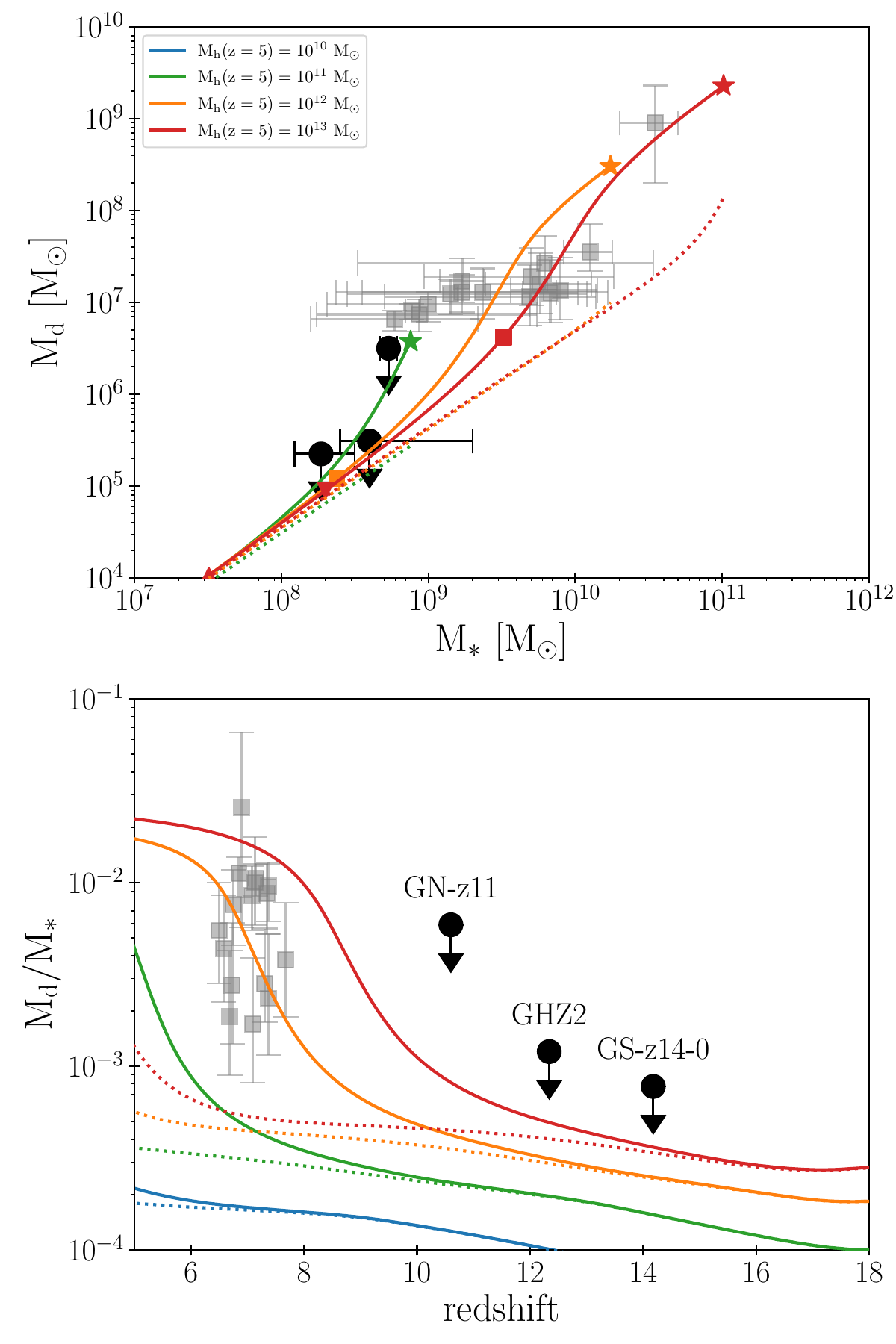}
\caption{
The dust mass evolution obtained from our model calculations with varying $M_{\rm h}(z = 5)$.
{\it Upper panel:} The $M_\ast$--$M_{\rm d}$ relation.
{\it Lower panel:} The redshift evolution of the dust-to-stellar mass ratio, $M_{\rm d}/M_\ast$.
Solid curves correspond to results with a metal accretion timescale of $\tau_0 = 5~\Myr$, 
while dotted curves represent $\tau_0 = 50~\Myr$. 
Colored markers indicate redshifts $z = 18,~15,~13,~10$, and $5$, represented by circles, triangles, inverted triangles, squares, and stars, respectively.
For comparison, we include observational data for $z \sim 7$ galaxies as gray squares \citep[][]{Hashimoto2019PASJ, Reuter2020ApJ, Bakx2021MNRAS, Sommovigo2022MNRAS}.
Additionally, upper-limits on the dust mass for the $z > 10$ galaxies, 
GN-z11 \citep{Fudamoto2024MNRAS}, 
GHZ2 \citep{Mitsuhashi2025arXiv}, 
and GS-z14-0 \citep{Carniani2024arXiv}, are shown as black circles.
These upper-limits are derived from the non-detection of dust continuum radiation, assuming a dust temperature of 80~K as analyzed in the cited studies.
}
\label{fig:Md_z}
\end{figure}
%%%%%%%%%%%%%%%%%%%%%%%%%%%%%%%%%%

%%%%%%%%%% fig:Td_z %%%%%%%%%%%%%%%%
\begin{figure}
\centering
\includegraphics[width=0.9\columnwidth]{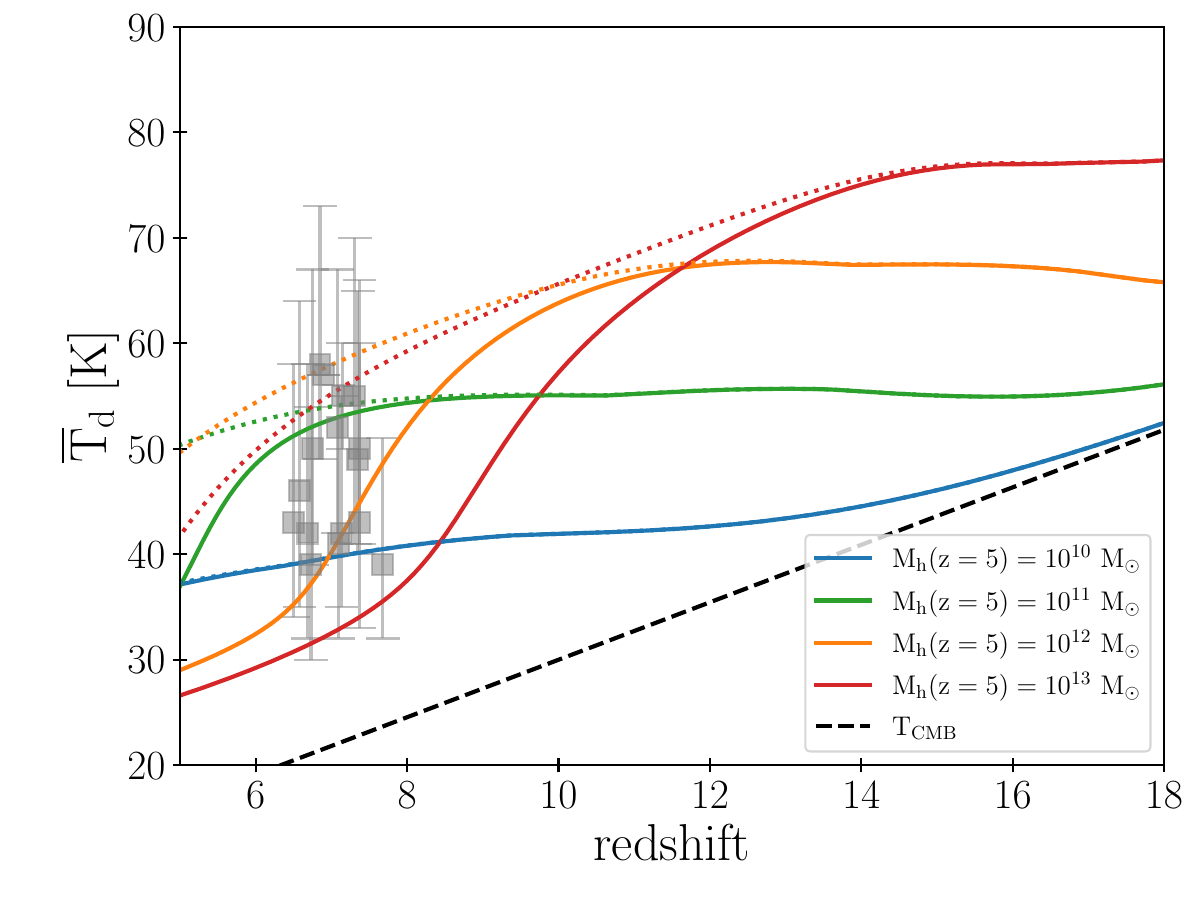}
\caption{
The flux-weighted dust temperature of our model galaxies as a function of redshift (see Eq.~(\ref{eq:Td_lw}) for the definition). 
Different colors and line styles correspond to those in Figure~\ref{fig:Md_z}, and gray squares represent the same observational data as in that figure.
For comparison, the CMB temperature at any redshift is shown with the black dashed line.
}
\label{fig:Td_z}
\end{figure}
%%%%%%%%%%%%%%%%%%%%%%%%%%%%%%%%%%

%%%%%%%%%% fig:FIR %%%%%%%%%%%%%%%%
\begin{figure}
\centering
\includegraphics[width=0.9\columnwidth]{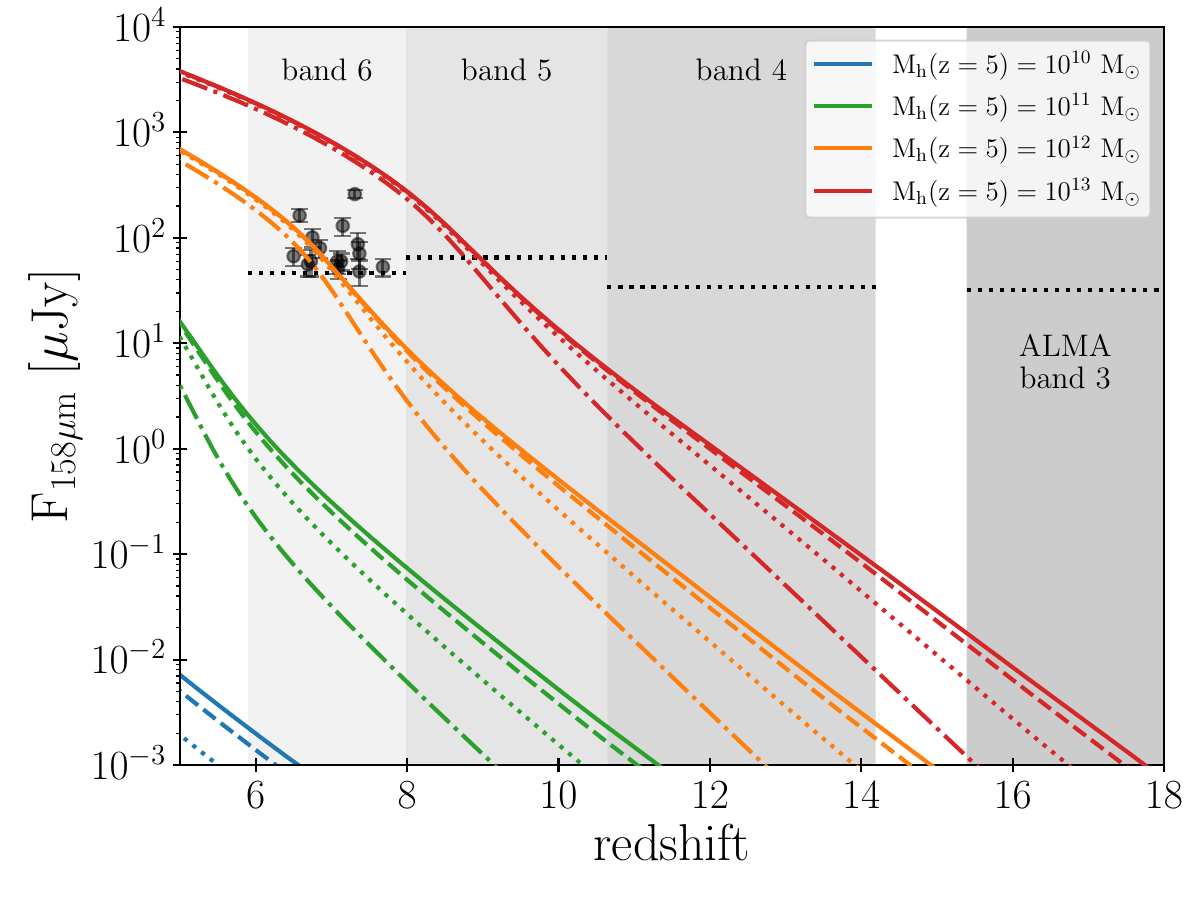}
\caption{
The redshift evolution of 158~$\micron$ dust continuum fluxes estimated for our model galaxies with varying $M_{\rm h}(z=5)$ is shown as the colored solid curves.
For comparison, the observed 158~$\micron$ fluxes for $z \sim 7$ galaxies (references indicated in Figure~\ref{fig:Md_z}) are shown as the black circles, while the 3$\sigma$ detection limits for one-hour ALMA exposures in Band 3-6 are represented by dotted horizontal lines.
This comparison demonstrates that our model calculations self-consistently explain both the presence of FIR-bright galaxies at $z \sim 7$ and the non-detection of dust continuum in UV-bright galaxies at $z > 10$.
Additionally, the colored dashed, dotted, and dash-dotted curves correspond to fluxes calculated assuming the optical depths reduced by factors of 0.5, 0.1, and 0.01, respectively, as also examined in Figure~\ref{fig:LF}.
}
\label{fig:FIR}
\end{figure}
%%%%%%%%%%%%%%%%%%%%%%%%%%%%%%%%%%

\subsection{Size evolution}\label{sec:M-size}

First, we explore the redshift evolution of the UV half-light radius $\reff$, which is defined as $\luv(< \reff) = 0.5~\luv(< \rmax)$ with Eq.~(\ref{eq:luv}).
In this analysis, we evaluate $\reff$ using the intrinsic UV luminosity rather than the dust-attenuated one, motivated by the marginal dust extinction observed in $z > 5$ galaxies. 
Potential explanations for this minimal dust attenuation are discussed in \S~\ref{sec:dust}.

Figure~\ref{fig:Reff} compares $\reff$ predicted by our model with those derived from the HST and JWST.
The upper and lower panels show how $\reff$ depends on $M_{\rm h}$ and $\fspin$, respectively.
We note that other parameters have a relatively minor impact on the size evolution.
Our model demonstrates that more massive galaxies are spatially more extended, with their sizes gradually increasing toward lower redshifts, which is generally consistent with the observational trend. 
The evolution of $\reff$ is attributed to our assumption that the spatial extent of galactic gas inflows scales with the virial radius, i.e., $h_{\rm R} \propto M_{\rm h}^{1/3} (1+z)^{-1}$.
This indicates that the observed size evolution of $z > 5$ galaxies is a natural consequence of structure formation in the $\Lambda$CDM cosmology.

However, we note discrepancies between our model predictions and observed correlations between ${\rm M_{UV}}$ and $\reff$.
Specifically, our calculations, fixing $\fspin = 0.5$ (upper panel), suggest that compact galaxies with $\reff < 100~\pc$ at $z \gtrsim 10$ are fainter than ${\rm M_{UV}} \sim -15~{\rm mag}$.
This prediction cannot explain the presence of very compact and bright galaxies, such as GN-z11 and GHZ2, which have $\reff \lesssim 100~\pc$ and ${\rm M_{UV}} < -20~{\rm mag}$.
Indeed, GN-z11 and GHZ2 may host active galactic nuclei (AGN), thereby appearing as compact, UV-bright sources.
However, the relative contributions of AGN and star formation to their UV luminosity remain highly uncertain \citep[][]{Maiolino2024Natur, Castellano2024ApJ}.
If star formation dominates their UV luminosity, a possible solution to this discrepancy is a diversity in $\fspin$ among high-$z$ galaxies.

As shown in the lower panel of Figure~\ref{fig:Reff}, varying $\fspin$ from 0.1 to 2 for halos with $M_{\rm h}(z=5) = 10^{12}~\msun$ generally reproduces the wide range of $\reff$ observed for $z > 10$ galaxies.
In particular, the case of $\fspin = 0.1$ produces GN-z11 and GHZ2-like galaxies around $z \sim 10$.
This result suggests that such extremely compact galaxies in the early universe may have experienced efficient angular momentum extraction, potentially due to disk instabilities \citep[][]{Noguchi1999ApJ, Immeli2004A&A, Bournaud2007ApJ} and galaxy mergers \citep[][]{Hernquist1989Natur, Barnes1991ApJ, Barnes1996ApJ, Hopkins2010MNRAS}.

\subsection{Mass-SFR-Metallicity relation}\label{sec:M-SFR-Z}

Figure~\ref{fig:ms_sfr_z} compares our calculated stellar mass-SFR-metallicity relation with corresponding observational data at different redshifts.
From top to bottom, we show the results for varying $M_{\rm h}(z=5)$, $\fboost$, $\fspin$, and $\fout$.
Overall, our model predictions are consistent with the observational data.

The left column of Figure~\ref{fig:ms_sfr_z} presents the $M_\ast$--SFR relation, showing that our model naturally produces SFRs about 10-100 times higher than those for local galaxies with comparable stellar masses.
We also find that SFRs increase with higher $M_{\rm h}$ and lower $\fout$, suggesting that the observed diversity in SFRs primarily reflects differences in halo mass growth histories and outflow strengths.
In contrast, variations in $\fboost$ and $\fspin$ have little impact on the evolutionary track in the $M_\ast$-SFR plane, 
even though higher $\fboost$ and lower $\fspin$, which increases $\sgmg$, likely enhance local star formation rates, $\sgmsf$.
As we will demonstrate in \S~\ref{sec:SFE}, this independence of global SFRs from the local star formation law and galaxy sizes naturally arises from a self-regulated evolution: an elevated $\sgmsf$ rapidly depletes gas within galaxies, preventing a sustained increase in the global SFRs.
Interestingly, this self-regulation process leads to a tight correlation between SFRs and $\dot{M}_{\rm h}$, suggesting that the enhanced SFRs in $z > 5$ galaxies primarily result from the large cosmological accretion rates in the early phase of galaxy evolution.

It is worth noting that our model predicts higher specific star formation rates $\sfr/M_\ast$ for lower-mass halos and at higher redshifts.
Consequently, galaxies with $M_{\rm h} \lesssim 10^{12}~\msun$ maintain $\sfr/M_\ast \gtrsim 20~{\rm Gyr^{-1}}$ until $z \sim 10$.
This satisfies the theoretical criterion for radiation pressure on dust grains to expel gas from galaxies \citep[][]{Fiore2023ApJ, Ferrara24}.
Such radiation-driven dusty outflows may play a crucial role in explaining the observed number density of UV-bright galaxies at $z > 10$, as we discuss in \S~\ref{sec:DA1}.

The middle column of Figure~\ref{fig:ms_sfr_z} shows the mass-metallicity relation,
revealing a strong dependence of gas metallicity on $\fout$, while other parameters have relatively minor effects.
Higher values of $\fout$ lead to lower metallicity due to the efficient evacuation of metals by outflows.
Our models with $\fout = 5$ and 20 closely match the median metallicity of the observational data,
whereas models with $\fout = 1$ and $100$ correspond to the upper and lower edges of the observed metallicity distribution, respectively.
This agreement suggests that the mass loading factor varies among high-$z$ galaxies by a few orders of magnitude, with an average value of $\fout \sim 10$.

We also examine the fundamental metallicity relation, which $z \lesssim 3$ star-forming galaxies are known to follow. 
This relation is given by,
\begin{eqnarray}
\oh = 0.43 \mu + 4.58 \ ,
\label{eq:FMR}
\end{eqnarray}
where $\mu = {\rm log}(M_\ast)-0.66{\rm log}({\rm SFR})$ \citep[e.g.,][]{Mannucci2010MNRAS, Andrews2013ApJ}.
Recent JWST observations of $z > 3$ galaxies indicate that their metallicities either follow the empirical law or are lower by a factor of a few \citep[][]{Nakajima2023ApJS, Curti2024A&A, Sarkar2024arXiv, Bunker2023AA, Arrabal_Haro2023aApJ, Castellano2024ApJ, D'Eugenio2024, Carniani2024arXiv}.
The right column of Figure~\ref{fig:ms_sfr_z} compares these observational results with our model calculations, 
showing that the observed trend can be explained by supposing high mass loading factors ($\fout \gtrsim 5$).
This indicates that powerful outflows are prevalent in $z > 5$ galaxies, significantly impeding the progress of chemical enrichment.

\subsection{Dust properties}\label{sec:Mdust}

Next, we explore the dust mass of galaxies ($M_{\rm d}$).
Figure~\ref{fig:Md_z} shows our calculation results with the fiducial case for the $M_\ast$--$M_{\rm d}$ relation (upper panel) and the redshift evolution of $M_{\rm d}/M_\ast$ (lower panel).
Our calculations are generally in good agreement with observations, successfully reproducing the dust content of $z \sim 7$ galaxies and remaining consistent with the upper limits for the galaxies at $z > 10$ (GN-z11, GHZ2, and GS-z14-0).
We suggest that dark matter halos growing to $M_{\rm h}(z=5) \gtrsim 10^{12}~\msun$ can represent dust-rich galaxies with $M_{\rm d} \gtrsim 10^7~\msun$ observed at $z \sim 7$.

A notable feature of our results is the rapid increase of $M_{\rm d}/M_\ast$ at $z < 10$, which is driven by dust growth via metal accretion.
In our fiducial case ($\tau_0 = 5~\Myr$), massive halos with $M_{\rm h}(z=5) \gtrsim 10^{12}~\msun$ achieve a short metal accretion timescale $\tau_{\rm acc} \lesssim 50~\Myr$ at $z \lesssim 10$ since their metallicities have reached $Z \sim 0.1~\zsun$.
As a result, rapid metal accretion causes $M_{\rm d}/M_\ast$ to increase by an order of magnitude by $z = 5$.

To explore the impact of slower dust growth, we perform additional calculations assuming a longer timescale ($\tau_0 = 50~\Myr$), shown as dotted curves in Figure~\ref{fig:Md_z}.
In this case, $M_{\rm d}/M_\ast$ shows only marginal growth even after $z \sim 10$, highlighting the importance of efficient metal accretion for explaining the observed dust enrichment in $z \sim 7$ galaxies.

We also investigate the dust temperature of our model galaxies.
With Eq.~(\ref{eq:Td}), we evaluate the dust temperature at any radii within the galaxies.
For convenience, we average the dust temperature across the galaxy as follows:
\begin{eqnarray}
\overline{T}_{\rm d} \equiv \frac{4 \pi \int T^\prime_{\rm d} I_{158} R{\rm d}R}{F_{158}} \ ,
\label{eq:Td_lw}
\end{eqnarray}
where the weighting is based on the dust continuum flux at 158~$\micron$.
This flux-weighted temperature is useful for comparing with the observed dust temperature of $z \sim 7$ galaxies, which are usually derived under the assumption of single blackbody radiation.

Figure~\ref{fig:Td_z} shows the redshift evolution of $\overline{T}_{\rm d}$ of our model galaxies.
The galaxy hosted by the least massive halo with $M_{\rm h}(z = 5) \sim 10^{10}~\msun$ has cold dust with $\overline{T}_{\rm d} \sim 40~\kelvin$ at $z \sim 10$, and the temperature remains nearly constant down to $z = 5$.
In contrast, galaxies hosted by more massive halos with $M_{\rm h}(z = 5) \gtrsim 10^{12}~\msun$ exhibit higher dust temperature with $\overline{T}_{\rm d} \sim 65~\kelvin$ at $z \sim 10$.
This hot dust is a consequence of high SFRs coupled with low dust masses, as $\overline{T}_{\rm d}$ approximately scales with $(\sfr/M_{\rm d})^{1/(4+\beta_{\rm d})}$ for optically thick galactic disks, as described by Eq.~(\ref{eq:Td}).

As redshift decreases, the dust temperature in massive galaxies declines rapidly, reaching $\overline{T}_{\rm d} \sim 30~\kelvin$ by $z \sim 5$. 
This decline is primarily driven by a reduction in $\sfr/M_{\rm d}$ due to efficient dust growth, as seen in Figure~\ref{fig:Md_z}.
This is evident from the fact that a longer metal accretion timescale $\tau_0 = 50~\Myr$ results in a more gradual decrease in dust temperature, as indicated by the dotted curves in Figure~\ref{fig:Td_z}.
Notably, the predicted $\overline{T}_{\rm d}$ values of our model galaxies are roughly consistent with the observed dust temperature of $z \sim 7$ galaxies.
This consistency suggests that, in the observed galaxies, dust heating by UV radiation becomes less effective as dust mass increases rapidly through metal accretion.

\subsubsection{Dust emission}
Finally, we discuss the detectability of dust continuum radiation from $z > 5$ galaxies.
Figure~\ref{fig:FIR} compares the 158~$\micron$ dust continuum fluxes predicted by our model with observed fluxes for $z \sim 7$ galaxies (black circles) and the 3$\sigma$ detection limits for one-hour ALMA exposure in Band 3-6 (dotted horizontal lines).
Our results show that dust continuum from galaxies hosted by halos with $M_{\rm h}(z=5) \sim 10^{12}~(10^{13})~\msun$ is detectable at $z \lesssim 6~(8)$.
This implies that FIR-bright galaxies observed at $z \sim 7$ are likely associated with such massive dark-matter halos.
Conversely, at $z > 10$, the predicted fluxes from our model galaxies are too faint for detection in the FIR band, 
consistent with the non-detection of dust continuum in UV-bright galaxies at $z > 10$, such as GN-z11 \citep{Fudamoto2024MNRAS}, GHZ2 \citep{Zavala2024arXiv, Mitsuhashi2025arXiv}, and GS-z14-0 \citep{Carniani2024arXiv, Schouws2024arXiv}.

The non-detectability of FIR emission at $z > 10$ has also been predicted by \citet{Ferrara2024arXiv}. 
They proposed that massive galaxies at $z > 10$ are likely to emit hot dust emission with temperatures $T_{\rm d} \gtrsim 70~\kelvin$ and a peak wavelength of $\lambda_{\rm peak} \lesssim 40~\mu{\rm m}$. 
As a result, the flux at $158~\mu{\rm m}$ is significantly reduced by a factor of $(\lambda_{\rm peak}/158~\mu{\rm m})^4$, being undetectable in the ALMA observations.
Our detailed model calculations successfully confirm this prediction, offering further evidence for the challenges of observing dust continuum emission from galaxies at $z > 10$.

%%%%%%%%%% fig:SFE %%%%%%%%%%%%%%%%
\begin{figure}
\centering
\includegraphics[width=0.9\columnwidth]{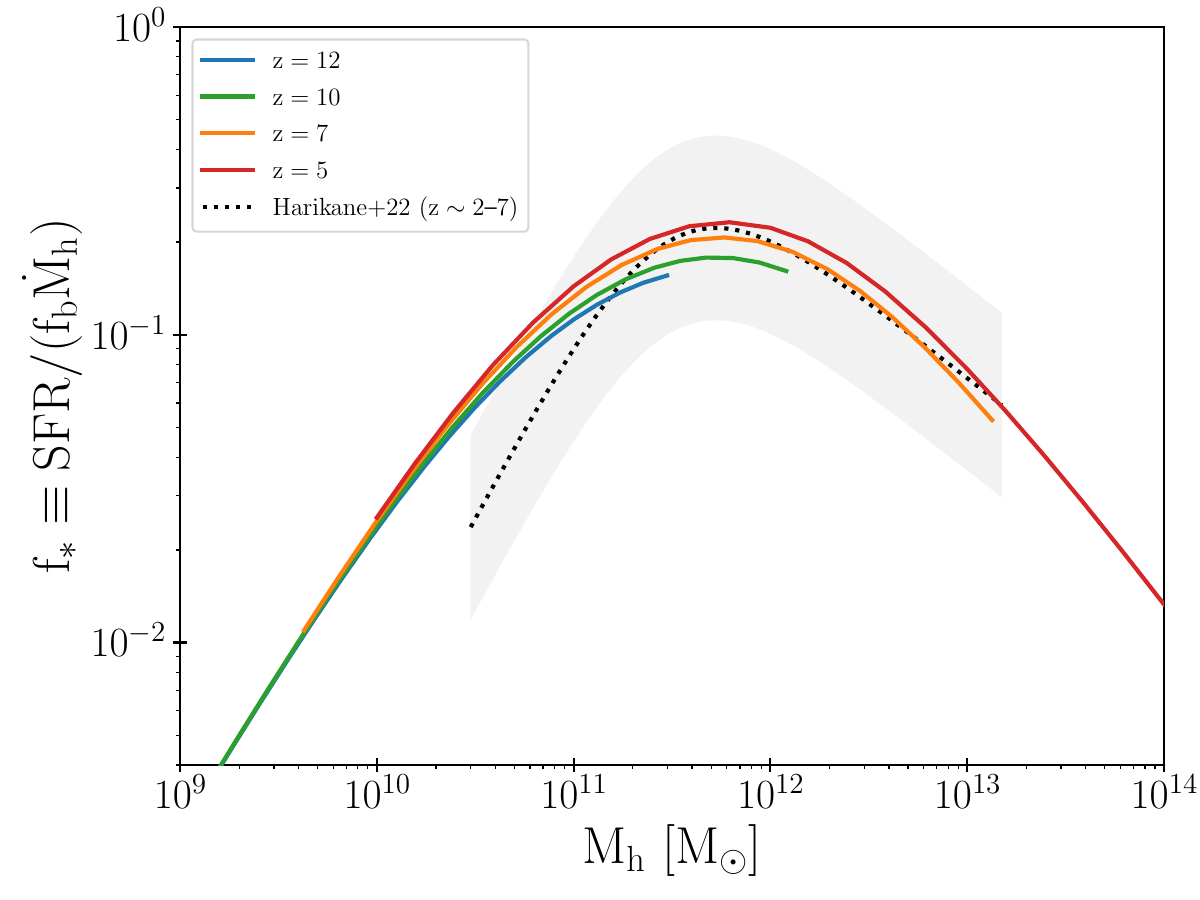}
\caption{
The $f_\ast$--$M_{\rm h}$ relation predicted by our fiducial model.
The blue, green, orange, and red curves correspond to the results at $z = 12$, 10, 7, and 5, respectively.
For comparison, the black dotted curve represents the fitting formula for the observed $f_\ast$--$M_{\rm h}$ relation of $z \sim 2$--$7$ galaxies, and the gray shaded region indicates the $2\sigma$ scatter in the observational data \citep[][]{Harikane2022ApJS}.
}
\label{fig:SFE}
\end{figure}
%%%%%%%%%%%%%%%%%%%%%%%%%%%%%%%%%%

%%%%%%%%%% fig:SFE_compare %%%%%%%%%%%%%%%%
\begin{figure*}
\centering
\includegraphics[width=\textwidth]{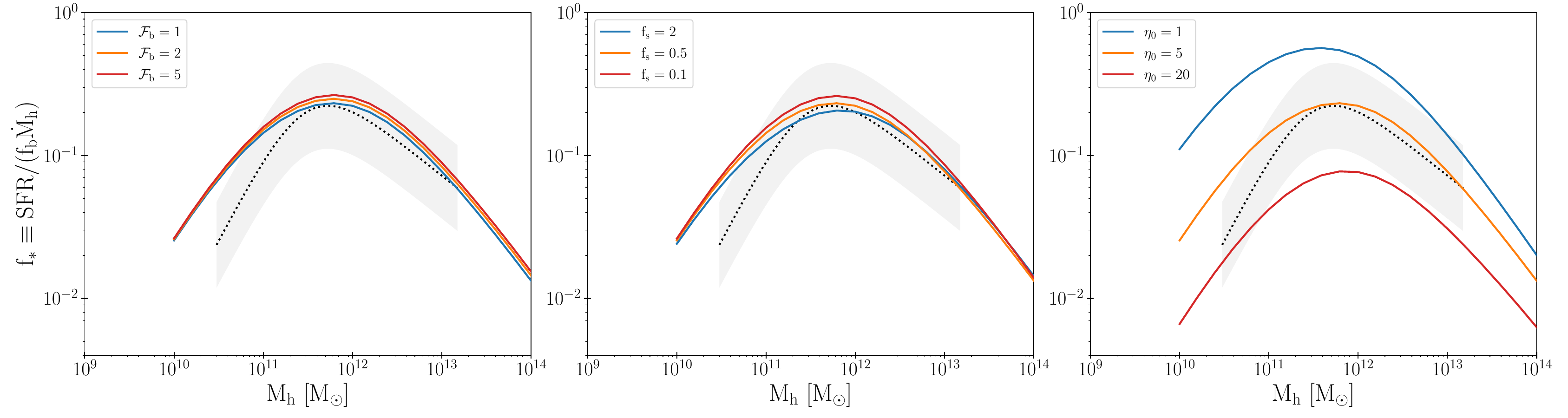}
\caption{
The $f_\ast$--$M_{\rm h}$ relation at $z = 5$, obtained from our calculations.
From left to right, the colored curves in each panel represent results with varying $\fboost$, $\fspin$, and $\fout$, with their specific values noted in the top-left corner of each left panel. 
The black dotted curve and the gray shaded region are the same as those shown in Figure~\ref{fig:SFE}.
}
\label{fig:SFE_compare}
\end{figure*}
%%%%%%%%%%%%%%%%%%%%%%%%%%%%%%%%%%

%%%%%%%%%% fig:mdot_Mh %%%%%%%%%%%%%%%%
\begin{figure}
\centering
\includegraphics[width=0.9\columnwidth]{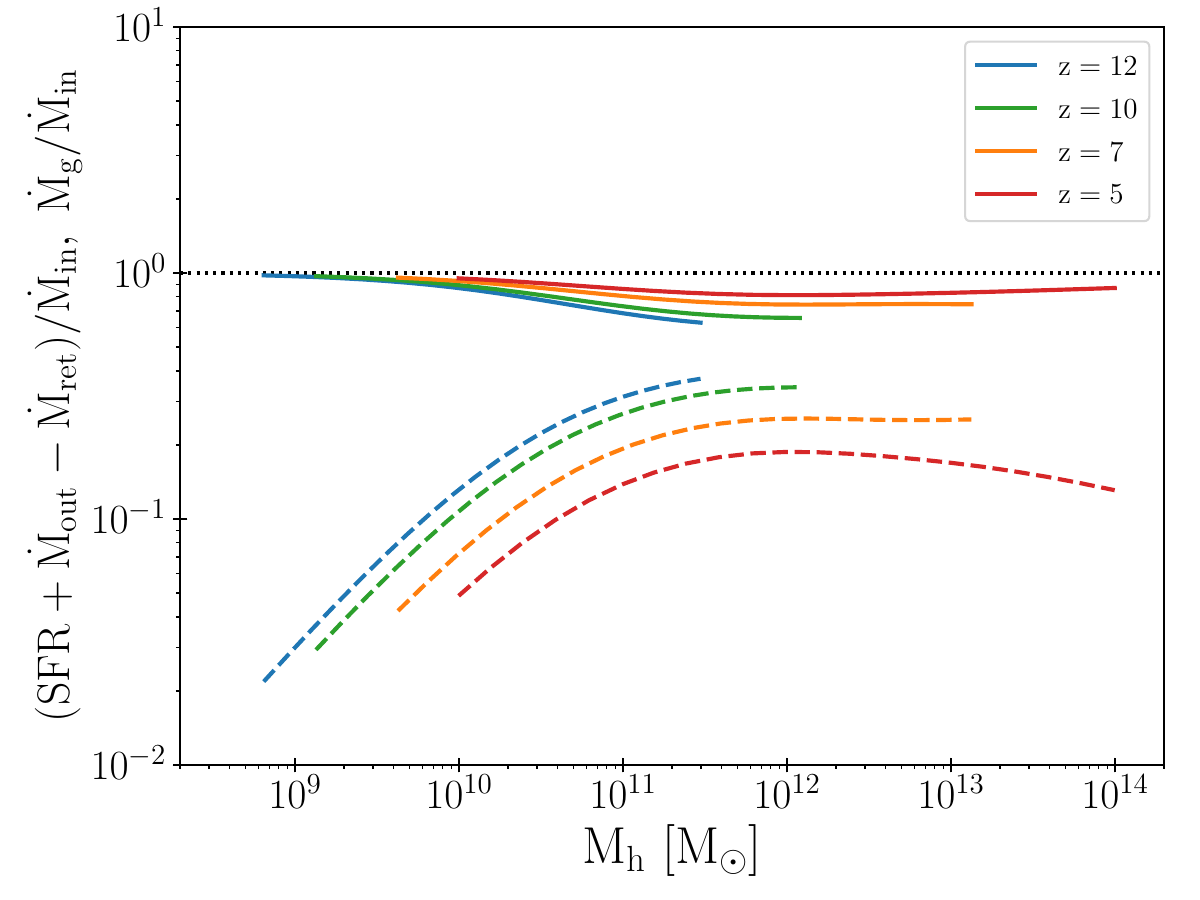}
\caption{
The comparison of the five terms in Eq.~(\ref{eq:cons_g_tot}) as a function of $z$ and $M_{\rm h}$, obtained in our fiducial model.
The solid curves represent the ratio of the gas consumption rate driven by star formation and outflows to the gas inflow rate, $(\sfr+\dot{M}_{\rm out}-\dot{M}_{\rm ret})/\dot{M}_{\rm in}$.
The dashed curves represent the ratio of the differential rate of gas mass to the gas inflow rate, $\dot{M}_{\rm g}/\dot{M}_{\rm in}$, respectively.
These ratios are shown for $z = 12$, 10, 7, and 5, with the blue, green, orange, and red curves, respectively.
}
\label{fig:mdot_Mh}
\end{figure}
%%%%%%%%%%%%%%%%%%%%%%%%%%%%%%%%%%

\section{Star formation efficiency}\label{sec:result2}

\subsection{Star formation efficiency at different halo masses}\label{sec:SFE}

As shown in \S~\ref{sec:results}, our fiducial model with $\fboost = 1$, $\fspin = 0.5$, $\fout = 5$, and $\tau_0 = 5~\Myr$ successfully explains various properties of $z > 5$ galaxies, including sizes, stellar masses, SFRs, metallicities, and dust contents.
These calculations allow us to meaningfully examine star formation activities for high-$z$ galaxies.

We characterize the star formation efficiency (SFE) as $f_\ast \equiv \sfr/(f_{\rm b}\dot{M}_{\rm h})$ following previous studies \citep[e.g.,][]{Harikane2018PASJ, Harikane2022ApJS, Inayoshi2022ApJ}. 
Figure~\ref{fig:SFE} shows the $f_\ast$--$M_{\rm h}$ relation at $z \geq 5$, derived from our fiducial model.
We find that $f_\ast$ follows an upward convex trend with $M_{\rm h}$, peaking around $M_{\rm h} \sim 10^{12}~\msun$,
and the peak value is $f_\ast \sim 0.2$ almost independent of redshift.
Our theoretical prediction is in good agreement with the empirical relation for $z \sim 2$--$7$ galaxies reported by \citet{Harikane2022ApJS}.
Recently, \citet{Donnan2025arXiv} have claimed that such a non-evolving SFE is essential to explain the observed UV LFs at $z \sim 6$-13 and the cosmic star formation rates at $z \sim 6$--$8$. 
These facts suggest that the $f_\ast$--$M_{\rm h}$ relation remains constant across a fairly wide redshift range.

Additionally, we examine how $f_\ast$ depends on the three parameters $\fboost$, $\fspin$, and $\fout$ in Figure~\ref{fig:SFE_compare}.
We find that while higher $\fboost$ and lower $\fspin$ results in slightly higher $f_\ast$ at any given $M_{\rm h}$, the variations are minimal.
This trend is consistent with what we observed in the $M_\ast$--SFR relation in Figure~\ref{fig:ms_sfr_z}.
In contrast, $f_\ast$ exhibits a much stronger dependence on $\fout$, increasing in inverse proportion to $\fout$,
while the overall shape of the $f_\ast$--$M_{\rm h}$ relation remains unchanged.
These results suggest that the SFE within a galaxy is predominantly governed by the total mass lost via outflows, rather than by the local star formation law or the spatial distribution of gas.

These trends in the $f_\ast$--$M_{\rm h}$ relation can be understood as follows.
By integrating Eq.~(\ref{eq:cons_g}) over the entire galactic disk, the time evolution of the total gas mass is given by,
\begin{eqnarray}
\dot{M}_{\rm g} 
&=& -(\sfr + \dot{M}_{\rm out}) + \dot{M}_{\rm in} + \dot{M}_{\rm ret} \nonumber \\
&=& -(1-\mathcal{R}+\eta) \sfr + f_{\rm b} \epsilon_{\rm in} \dot{M}_{\rm h} \ ,
\label{eq:cons_g_tot}
\end{eqnarray}
where the first term in the second line represents net gas consumption due to star formation and outflows, 
and $\mathcal{R}$ denotes the return mass fraction from stellar evolution, approximated as $\mathcal{R} \sim 0.5$ in our numerical setup.
As shown in Figure~\ref{fig:mdot_Mh}, the gas consumption rate is comparable to the gas inflow rate, regardless of redshift and halo mass. 
As a result, one obtains $|\dot{M}_{\rm g}| \ll \dot{M}_{\rm in}$, implying that our model galaxies evolve in a quasi-steady state, where the gas consumption balances with the gas supply.

This equilibrium would be a natural outcome of a self-regulating mechanism.
When gas supply exceeds gas consumption in a galaxy, the increased gas mass triggers enhanced star formation and subsequent outflows, which in turn accelerates gas consumption.
Conversely, when gas consumption surpasses gas supply, the galaxy depletes its gas reservoir, leading to reduced SFRs.
As a result, galaxies naturally settle into an equilibrium state.
This self-regulation also explains the weak dependence of $f_\ast$ on $\fboost$ and $\fspin$.
Although higher $\fboost$ and lower $\fspin$ enhance local star formation rates $\sgmsf$,
this accelerates gas depletion within the galaxy, preventing a sustained increase in the global SFR.

Then, by assuming $\dot{M}_{\rm g} \sim 0$ in Eq.~(\ref{eq:cons_g_tot}), we can approximate $f_\ast$ as,
\begin{eqnarray}
f_\ast \equiv \frac{\sfr}{f_{\rm b} \dot{M}_{\rm h}} \sim \frac{\epsilon_{\rm in}}{1-\mathcal{R}+\eta} \ .
\label{eq:f_star}
\end{eqnarray}
This yields $f_\ast \sim 0.18$ for $M_{\rm h} = 10^{11}~\msun$ by assuming $\fout = 5$ as in our fiducial model.
This formula highlights that $f_\ast$ is characterized by $\epsilon_{\rm in}$ and $\eta$.
Specifically, our model predicts 
\begin{equation}
\hspace{4em} f_\ast \sim 
\begin{cases}
\eta^{-1} \propto M_{\rm h}^{1.1} ,  & (M_{\rm h} \ll 10^{11}~\msun) \\  
\epsilon_{\rm in}/(1-\mathcal{R}) \propto M_{\rm h}^{-1} &  (M_{\rm h} \gg 10^{12}~\msun) \ .
\end{cases}
\label{eq:f_star_limit}
\end{equation}

These trends reflects the underlying physics of viral shocks and supernova-driven outflows, as assumed in our functional forms of $\epsilon_{\rm in}$ and $\eta$ (see \S~\ref{sec:SF_IF_OF}).
Furthermore, if $\epsilon_{\rm in}$ and $\eta$ are nearly independent of redshifts,
our formulation naturally explains the weak redshift evolution of $f_\ast$ observed for $z < 7$ galaxies \citep[][]{Harikane2018PASJ, Harikane2022ApJS, Donnan2025arXiv}.
Conversely, if a higher $f_\ast$ is required to match cosmic SFR density at $z \gtrsim 10$, as suggested by \citet{Harikane2023ApJS},
this would imply a need to reduce $\eta$, given that further increasing $\epsilon_{\rm in}$ beyond unity is unphysical.
We will explore this point more explicitly in \S~\ref{sec:CSFRD} and \ref{sec:fbf}.

Thus, the $f_\ast$--$M_{\rm h}$ relation reflects the quasi-steady evolution of galaxies, thereby encapsulating the nature of galactic inflows and outflows.
It is worth noting here that the quasi-steady behavior in our model stems from the assumption of continuous dark matter halo assembly history given by Eq.~(\ref{eq:mdoth}).
In reality, galaxies temporarily deviate from the quasi-steady state, as episodic gas inflows driven by galaxy mergers lead to intermittent star formation and outflows.
Nevertheless, our results suggest that, over longer timescales, galaxies tend to evolve toward a balance between gas supply and consumption, resulting in a small statistical scatter and weak redshift dependence in the observed $f_\ast$--$M_{\rm h}$ relation.
To fully understand the impact of bursty star formation histories on the statistical properties of $f_\ast$,
it is crucial to extend our model to incorporate realistic mass assembly histories.

%%%%%%%%%% fig:LF %%%%%%%%%%%%%%%%
\begin{figure}
\centering
\includegraphics[width=0.9\columnwidth]{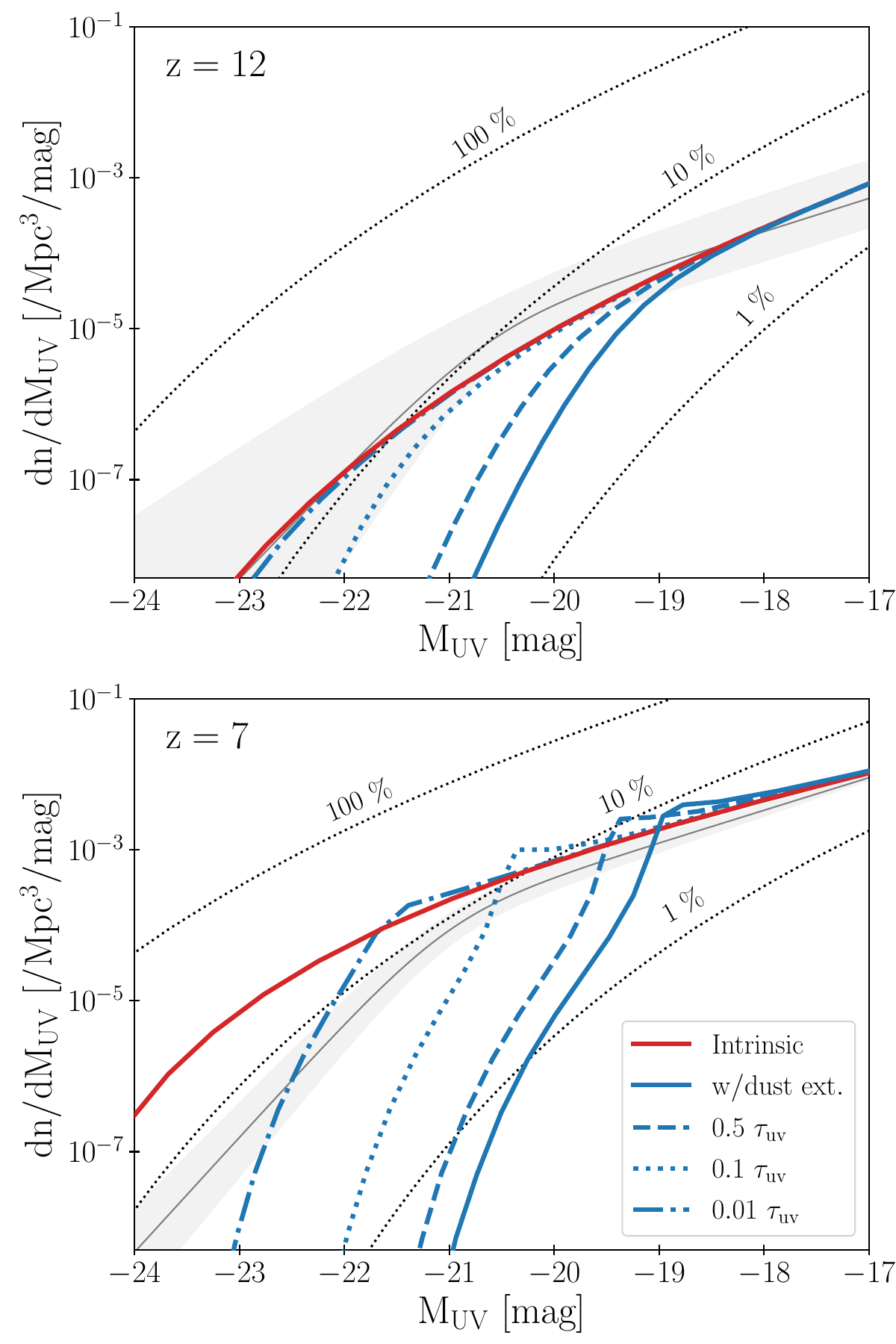}
\caption{
The UV LFs at $z = 12$ (upper panel) and $7$ (lower panel) derived from our model calculations.
The red solid curve represents the intrinsic UV LF, while the blue solid curve indicates the dust-attenuated UV LF.
To highlight the impact of dust extinction, dust-attenuated UV LFs with optical depths reduced by factors of 0.5, 0.1, and 0.01 are shown as the blue dashed, dotted, and dash-dotted curves, respectively.
The observed UV LFs and their $1\sigma$ uncertainties presented by \citet{Harikane2024barXiv} are depicted by the gray curves and shaded regions.
For comparison, the black dotted curves correspond to UV LFs predicted assuming constant SFEs of $f_\ast = 100~\%$, $10~\%$, and $1~\%$.
}
\label{fig:LF}
\end{figure}
%%%%%%%%%%%%%%%%%%%%%%%%%%%%%%%%%%

%%%%%%%%%% fig:tau %%%%%%%%%%%%%%%%
\begin{figure}
\centering
\includegraphics[width=0.9\columnwidth]{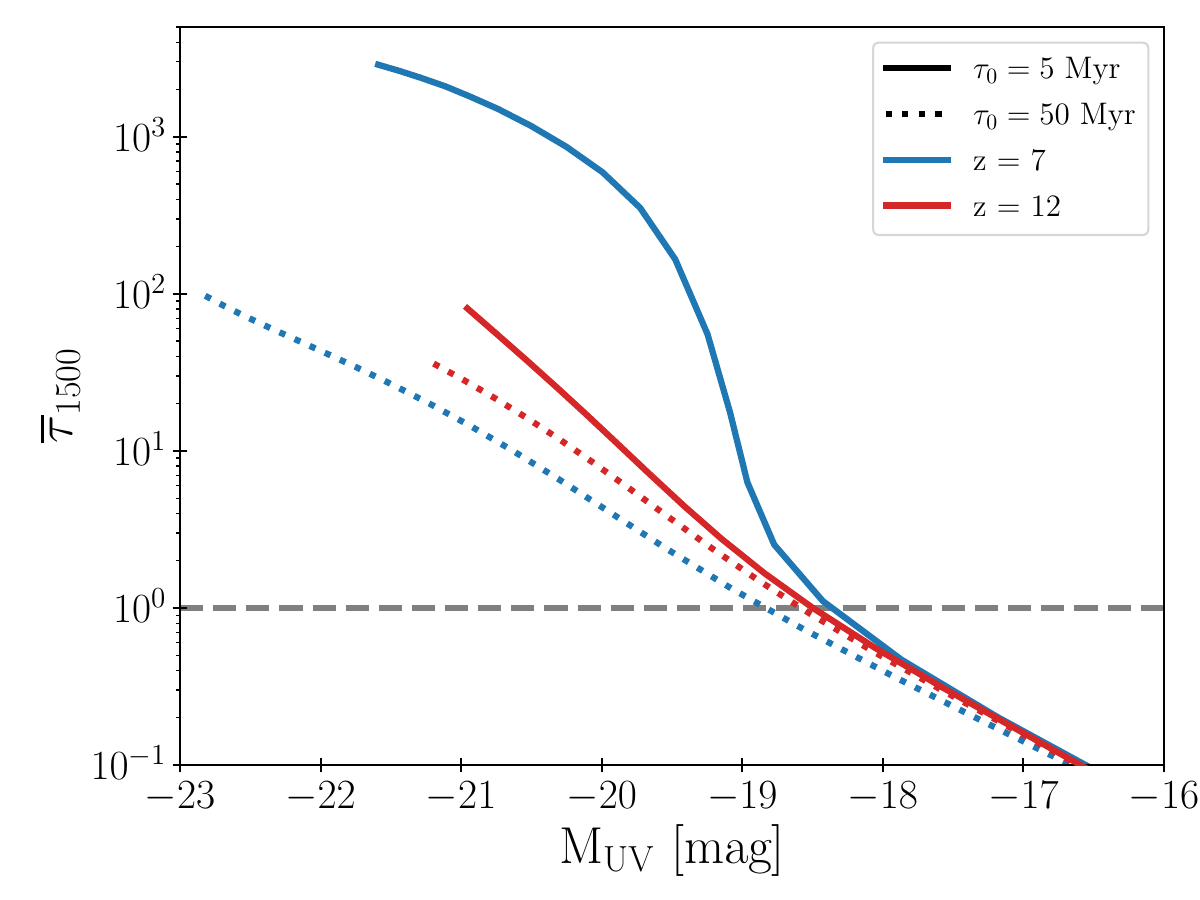}
\caption{
The luminosity-weighted dust optical depths, $\overline{\tau}_{1500}$, of our model galaxies as a function of dust-attenuated UV magnitude.
The red and blue curves represent results at $z = 12$ and 7, respectively.
The solid curves correspond to our fiducial model with a metal accretion timescale of $\tau_0 = 5~\Myr$, while dashed curves indicate results for the model with $\tau_0 = 50~\Myr$.
}
\label{fig:tau}
\end{figure}
%%%%%%%%%%%%%%%%%%%%%%%%%%%%%%%%%%

\subsection{UV luminosity functions and implications for dust attenuation}\label{sec:UV-LF}

Our model predicts the intrinsic and dust-attenuated UV magnitudes of galaxies as functions of host dark-matter halo masses and redshifts.
By combining these relations with the Sheth-Tormen dark-matter halo mass function \citep[][]{Sheth2002MNRAS}, 
we can construct UV LFs at any redshifts.

The upper panel of Figure~\ref{fig:LF} shows our theoretical UV LFs at $z = 12$ as well as an observed UV LF inferred with spectroscopically confirmed galaxy samples by \citet{Harikane2024barXiv}.
For comparison, we also show LFs calculated assuming constant SFEs of $f_\ast = 100\%$, $10\%$, and $1\%$.
Notably, our intrinsic LF (red curve) broadly agrees with the observed LF over the range from $M_{\rm UV} \sim -23~{\rm mag}$ to $-18~{\rm mag}$.
This agreement arises from the fact that the faint-end and bright-end galaxies are hosted by halos with $M_{\rm h} \sim 10^{10}~\msun$ and $10^{11}~\msun$
and form stars with SFEs of $f_\ast \sim 3~\%$ and $10~\%$, respectively, as shown in Figure~\ref{fig:SFE}.

We note, however, that the dust-attenuated LF (blue solid curve) shows significant suppression at $M_{\rm UV} \lesssim -20~{\rm mag}$ compared to the intrinsic one. 
This suppression is due to the high optical depths of these galaxies.
To quantify this effect, we evaluate the luminosity-weighted optical depth of each model galaxy as follows:
\begin{eqnarray}
\overline{\tau}_{1500} = \frac{4 \pi \int^{\rmax}_{\rmin} \tau_{1500} f_{\rm esc} \sgml R {\rm d}R}{L_{\rm UV}} \ .
\label{eq:tau_LW}
\end{eqnarray}
The red solid curve in Figure~\ref{fig:tau} shows the calculated $\overline{\tau}_{1500}$ at $z = 12$, revealing that galaxies with $M_{\rm UV} \lesssim -20~{\rm mag}$ exhibit optical depths $\overline{\tau}_{1500} \gtrsim 10$. 
These high optical depths result from high surface gas densities of $\sgmg \gtrsim 10^3~\msun \pc^{-2}$ with dust-to-gas mass ratios of $D/D_{\rm MW} \sim 0.1$, as indicated from Eq.~(\ref{eq:tau_uv}).
The emergence of highly obscured galaxies at $z \gtrsim 10$ has been highlighted in several studies \citep[e.g.,][]{Ziparo2023MNRAS, Ferrara2023MNRAS, Ferrara2024arXiv},
emphasizing the necessity of mechanisms to alleviate the impact of dust extinction.

To quantitatively address this, we examine how much the dust optical depth needs to be reduced to reproduce the observed UV LF.
The different blue curves in Figure~\ref{fig:LF} represent dust-attenuated LFs with $\tau_{1500}$ manually decreased by factors of 0.5, 0.1, and 0.01.
This analysis reveals that a $\gtrsim 90~\%$ reduction in $\tau_{1500}$ is necessary for our model to be consistent with observations.
Indeed, such low $\tau_{1500}$ has also been invoked to explain extremely blue UV colors of $z \gtrsim 10$ galaxies \citep[e.g.,][]{Topping2022ApJ, Topping2024MNRAS, Cullen2024MNRAS, Morales2024ApJ, Yanagisawa2024arXiv}.

Moreover, we conduct a similar analysis for the UV LFs at $z = 7$, as shown in the lower panel of Figure~\ref{fig:LF}.
The intrinsic UV LF at this redshift is approximately ten times higher than the observed one near the bright end, suggesting a need for non-negligible dust extinction to reconcile the discrepancy with observations.

On the other hand, the dust-attenuated UV LF (blue solid curve) falls significantly below the observed one at $M_{\rm UV} \lesssim -20~{\rm mag}$. 
The sharp decline is caused by rapid dust mass growth via metal accretion in these massive galaxies.
The blue curves in Figure~\ref{fig:tau} show $\overline{\tau}_{1500}$ of model galaxies at $z = 7$, obtained for different metal accretion timescales.
In our fiducial model with $\tau_{0} = 5~\Myr$, the optical depths increase dramatically at $M_{\rm UV} \sim -19~{\rm mag}$, reaching $\overline{\tau}_{1500} \sim 10^3$ for galaxies with $M_{\rm UV} \lesssim -20~{\rm mag}$.
In contrast, the slower metal accretion model with $\tau_{0} = 50~\Myr$ does not exhibit such a sharp rise in $\overline{\tau}_{1500}$.
Thus, the steep cutoff in our predicted dust-attenuated UV LF is a direct consequence of the short metal accretion timescale $\tau_{0} = 5~\Myr$, which is required to explain the dust content observed in $z \sim 7$ galaxies (see Figure~\ref{fig:Md_z}).
However, it is important to note that even the $\tau_{0} = 50~\Myr$ model underestimates the number density of bright-end galaxies compared to observations, as it still predicts $\overline{\tau}_{1500} \gtrsim 10$ for galaxies with $M_{\rm UV} \lesssim -21~{\rm mag}$.

Additionally, the different blue curves show that to explain the observed galaxy abundance at $M_{\rm UV} \sim -23~{\rm mag}$, $\tau_{1500}$ needs to be reduced by a factor of $\sim 0.01$.
However, even with such a significant reduction, this model fails to replicate the overall slope of the bright end of the observed UV LF.
Thus, at $z = 7$, the discrepancy between the model prediction and the observation is more pronounced than at $z = 12$, highlighting the critical importance of exploring dust physics across a wide redshift range.

We note here that if UV attenuation is suppressed by any mechanisms, it can reduce the energy absorbed by dust, thereby dimming the dust continuum radiation.
Consequently, the FIR fluxes derived using the original optical depths (solid curves in Figure~\ref{fig:FIR}) would be regarded as optimistic upper limits.
To demonstrate this, Figure~\ref{fig:FIR} also presents FIR fluxes assuming the reduced optical depths, shown by different line styles.
We find that a reduction in $\tau_{1500}$ up to $\sim 90~\%$ has minimal impact on FIR fluxes,
as the reduced optical depths remain large enough for substantial UV photon absorption.
In contrast, when $\tau_{1500}$ is reduced by $99~\%$, the FIR fluxes drop by approximately an order of magnitude at $z \gtrsim 10$.
This effect becomes less pronounced at lower redshifts, where the original optical depths are higher.
This analysis suggests that detecting the FIR dust continuum in $z > 10$ galaxies could be challenging, especially if UV photons escape more efficiently than anticipated by our model calculations.

\subsection{Cosmic star formation rate density}\label{sec:CSFRD}

In \S~\ref{sec:SFE}, we find that the $f_{\ast}$--$M_{\rm h}$ relation is generally preserved across a wide redshift range as a consequence of quasi-steady galaxy evolution.
We examine whether the non-evolving $f_{\ast}$--$M_{\rm h}$ relation is consistent with recent measurements of cosmic SFR densities (CSFRDs) for $z > 8$ galaxies.
CSFRDs are calculated as follows:
\begin{eqnarray}
\rho_{\rm SFR} 
&=& \int \sfr \frac{{\rm d}n}{{\rm d} M_{\rm UV}} {\rm d}M_{\rm UV} \nonumber \\
&=& \int f_\ast f_{\rm b} \dot{M}_{\rm h} \frac{{\rm d}n}{{\rm d} M_{\rm h}} {\rm d}M_{\rm h} \ ,
\label{eq:CSFRD}
\end{eqnarray}
where the integration is typically carried out down to $M_{\rm UV}^{\rm lim} = -17~{\rm mag}$ or up to the corresponding halo mass \citep[e.g.,][]{Madau2014ARA&A, Oesch2014ApJ, Finkelstein2015ApJ, McLeod2016MNRAS}.
\citet{Harikane2022ApJS} calculated Eq.~(\ref{eq:CSFRD}) under the assumption of a non-evolving $f_{\ast}$--$M_{\rm h}$ relation observed at $z \sim 2$-7 
and demonstrated that CSFRDs decline rapidly at $z \gtrsim 8$ following $\rho_{\rm SFR} \propto 10^{-0.5(1+z)}$.

\begin{figure}
\centering
\includegraphics[width=0.9\columnwidth]{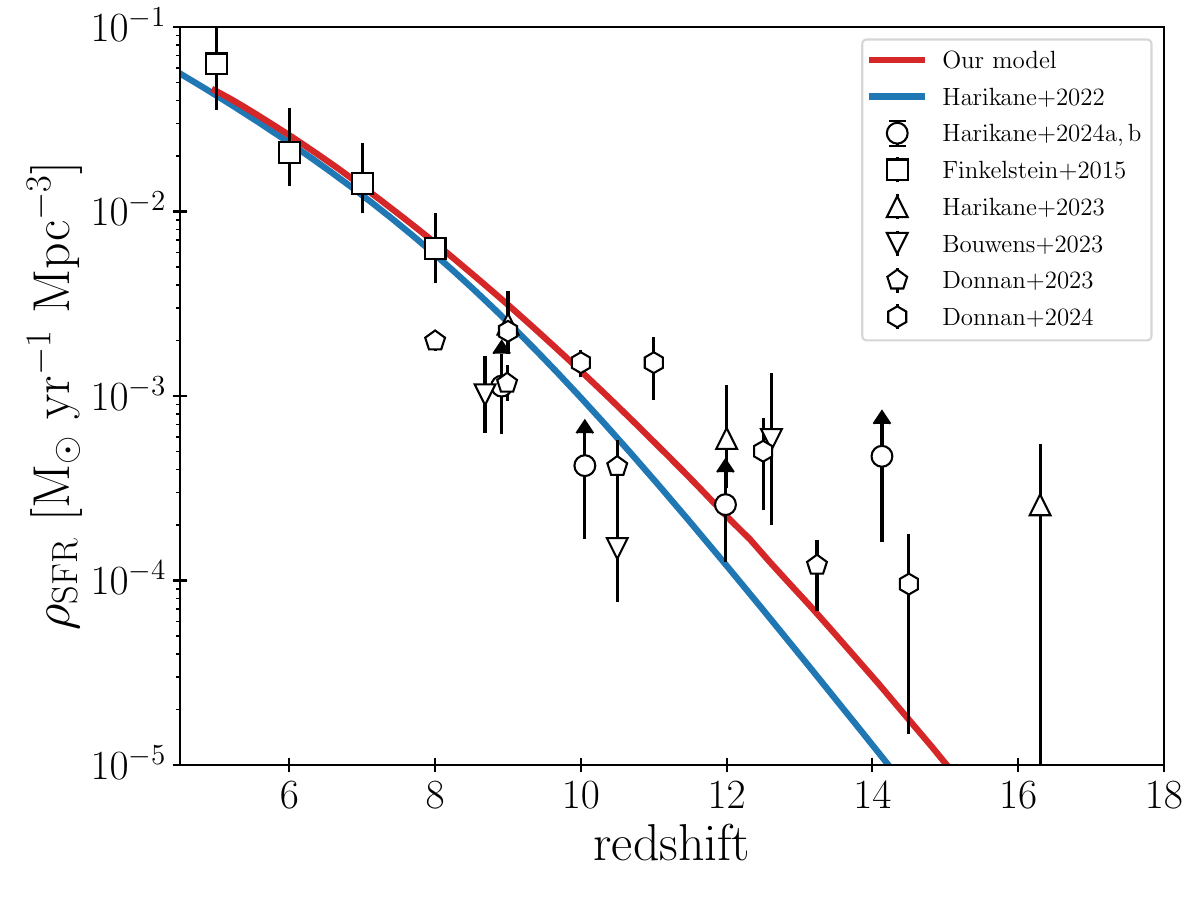}
\caption{
The cosmic star formation rate density obtained by our fiducial model is shown with the red curve.
The blue curve corresponds to the fitting formula to the observed CSFRDs in $z \lesssim 7$ presented by \citet{Harikane2022ApJS}.
For comparison, different open symbols represent observed cosmic star formation rate densities at $z \gtrsim 5$ \citep[][]{Finkelstein2015ApJ, Harikane2023ApJS, Bouwens2023MNRAS, Donnan2023MNRAS, Donnan2024MNRAS}.
}
\label{fig:CSFR}
\end{figure}

We evaluate CSFRDs using our calculation results.
In this analysis, we neglect the effects of dust attenuation as observed galaxies at $z > 8$ are nearly obscuration-free, as mentioned in \S~\ref{sec:UV-LF}.
Figure~\ref{fig:CSFR} compares our model-predicted CSFRDs with the theoretical result from \citet{Harikane2022ApJS} and recent observational measurements at $z \gtrsim 8$.
Our prediction is in good agreement with the model of \citet{Harikane2022ApJS},
which is expected since the $f_{\ast}$--$M_{\rm h}$ relation in our model is generally consistent with their results (see Figure~\ref{fig:SFE}).
Additionally, our model aligns well with recent observational data up to $z \sim 13$.
However, at $z \gtrsim 14$, our prediction tends to underestimate CSFRDs compared to observations,
implying that SFEs in the early universe must be a few times greater than the $f_{\ast}$--$M_{\rm h}$ relation predicted by our model.
One potential explanation for this discrepancy is the feedback-free galaxy evolution scenario, proposed by \citet{Dekel2023MNRAS}.
This scenario supposes that in extremely early galaxies, high gas density results in a short free-fall time for gas clouds, causing rapid star formation.
As a result, a large fraction of gas turns out stars before the onset of supernovae, and mass loss by galactic outflows can be negligible, i.e., $\eta \ll 1$.
According to Eq.~(\ref{eq:f_star}), such conditions yield a high SFE of $f_\ast \sim 1$.
Thus, if feedback-free galactic evolution is common at $z \gtrsim 14$, it could explain the high CSFRDs observed at these redshifts.
However, we claim that this extremely efficient star formation must cease before $z \sim 12$ to maintain the excellent agreement between our model predictions and observations across multiple properties, such as metallicities, UV LFs, and CSFRDs.
In \S~\ref{sec:fbf}, we explicitly examine the potential effects of feedback-free evolution on observed properties of high-$z$ galaxies.

%%%%%%%%%% fig:CSFR %%%%%%%%%%%%%%%%
%%%%%%%%%%%%%%%%%%%%%%%%%%%%%%%%%%

%%%%%%%%%%%%%%%%%%%%%%%%%%%%%%%%%%%%%%%%%%%%%%%%%%

\section{Discussion}\label{sec:discuss}

%%%%%%%%%% fig:LF_fc %%%%%%%%%%%%%%%%
\begin{figure}
\centering
\includegraphics[width=0.9\columnwidth]{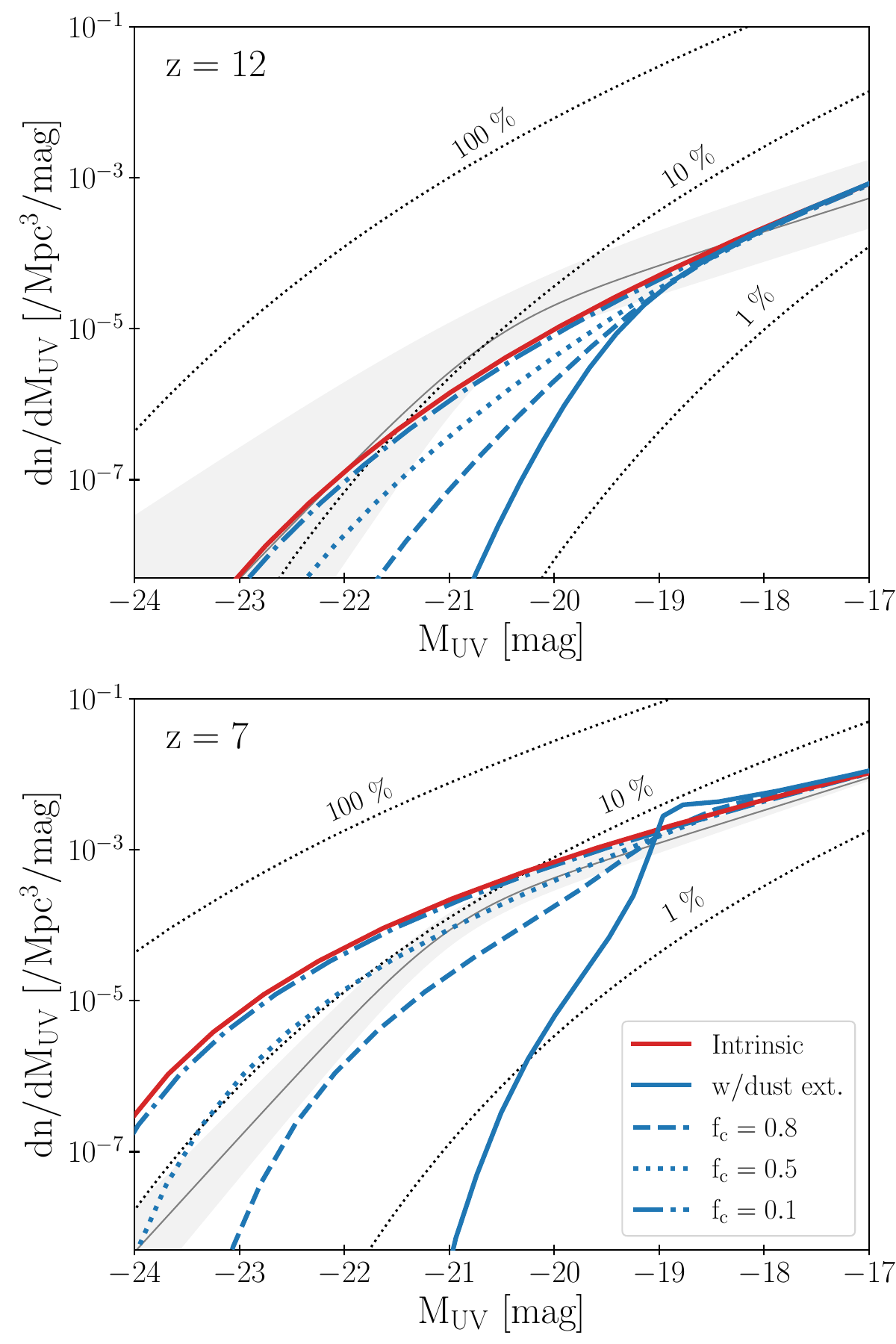}
\caption{
The same as Figure~\ref{fig:LF}, but showing UV LFs calculated for different values of $f_{\rm c}$ the dust covering fraction of young massive stars.
The results for $f_{\rm c} =$ 0.8, 0.5, and 0.1 are shown as the blue dashed, dotted, and dash-dotted curves, respectively.
The intrinsic and fully dust-attenuated UV LFs, represented by the red and blue solid curves, correspond to the extreme cases of $f_{\rm c} = 0$ and $1$, respectively.
}
\label{fig:LF_fc}
\end{figure}
%%%%%%%%%%%%%%%%%%%%%%%%%%%%%%%%%%

%%%%%%%%%% fig:ms_sfr_z_fbf %%%%%%%%%%%%%%%%
\begin{figure*}
\centering
\includegraphics[width=\textwidth]{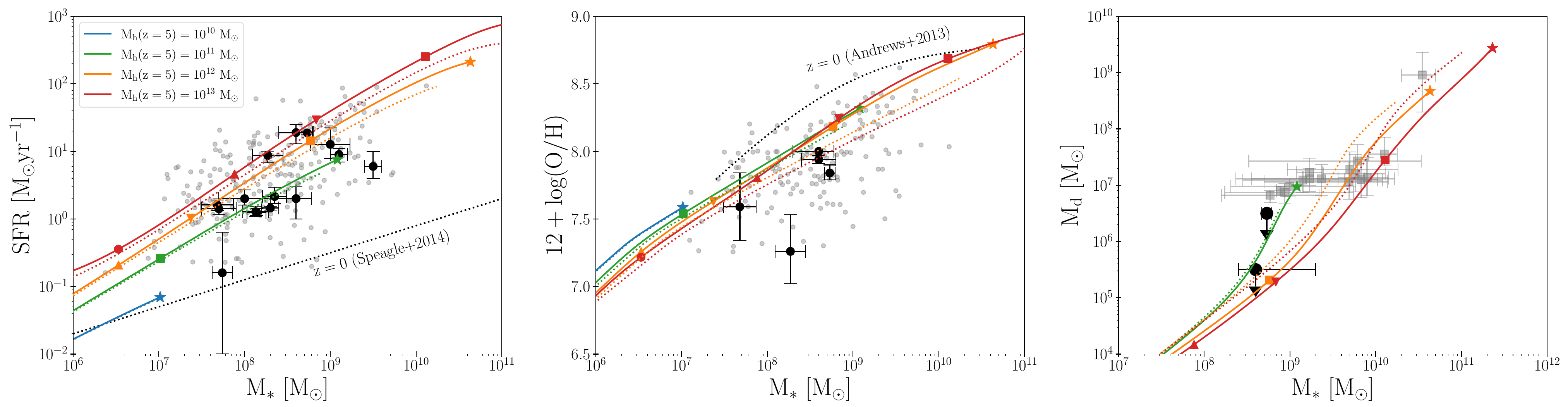}
\caption{
The left, middle, and right panels present the results obtained by the weak feedback model for the star formation rate ($\sfr$), gas metallicity ($\oh$), and dust mass ($M_{\rm d}$) as functions of the stellar mass ($M_\ast$), respectively.
The lines and markers carry the same meaning as those in Figures~\ref{fig:ms_sfr_z}~and~\ref{fig:Md_z}.
For comparison, the corresponding results from the fiducial model are shown as colored dotted curves.
}
\label{fig:ms_sfr_z_fbf}
\end{figure*}
%%%%%%%%%%%%%%%%%%%%%%%%%%%%%%%%%%

\subsection{How do galaxies avoid significant dust attenuation?}\label{sec:DA}

In~\S~\ref{sec:UV-LF}, we find that significant dust attenuation leads to a large discrepancy between our theoretical predictions and the observed UV LFs.
Here, we explore potential mechanisms that could mitigate the effects of dust attenuation without altering the fundamental characteristics of our model galaxies.
In the following discussion, we consider the one-zone averaged dust optical depth given by:
\begin{align}
&\tau_{1500} \sim \kappa_{1500} \frac{D M_{\rm g}}{2 \pi R_{\rm eff}^2} \nonumber \\
&\sim 34.6~
\left ( \frac{\kappa_{1500}}{1.26 \times 10^5~{\rm cm^2 g^{-1}}} \right ) 
\left ( \frac{M_{\rm g}}{10^{10}~\msun} \right ) 
\left ( \frac{R_{\rm eff}}{500~\pc} \right )^{-2}
\left ( \frac{D/D_{\rm MW}}{0.01} \right ) \ ,
\label{eq:tau_uv_gal}
\end{align}
where the reference values of $M_{\rm g}$, $R_{\rm eff}$, and $D$ are representative of bright-end galaxies at $z \sim 12$.
A straightforward approach to reduce $\tau_{1500}$ is to decrease either $D$ or $M_{\rm g}$.
Our model can achieve this by supposing more gas and dust evacuation from galaxies with higher values of $\fout$.
However, further increasing $\fout$ would lead to gas depletion in galaxies, 
making it difficult to reproduce the $M_\ast$--$\sfr$--$Z$ relation of $z > 5$ galaxies.
Furthermore, excessively reducing $D$ could conflict with the presence of dust-rich galaxies observed at $z \sim 7$.
Therefore, we disfavor this approach as an effective way to reduce $\tau_{1500}$.
Alternatively, we discuss three possibilities in the following.

\subsubsection{Dust displacement}\label{sec:DA1}
The first possibility is that dust is displaced to large ($\sim $kpc) scales by radiation-driven outflows.
\citet{Fiore2023ApJ} have proposed that in $z > 10$ galaxies, radiation pressure on dust grains is usually strong enough to drive substantial dusty outflows \citep[see also][]{Fukushima2018MNRAS, Ferrara2023MNRAS}.
While these dusty outflows are likely still gravitationally bound by the galaxy and expected to eventually return to the galactic disk, they can temporarily extend the spatial distribution of dusty gas relative to the stellar component \citep[][]{Ziparo2023MNRAS, Ferrara2024arXiv}.
If the effective radius of dusty gas ($R_{\rm eff,d}$) increases to $R_{\rm eff,d} \gtrsim 5~R_{\rm eff}$, $\tau_{1500}$ decreases by $\gtrsim 25$.
This reduction is sufficient to reconcile the discrepancy between the UV LF predicted by our model and the observed one at $z \sim 12$.

A potential concern with this scenario is that substantial dust removal from the main galaxy body shuts down star formation, leading to a decline in UV luminosity of the galaxy.
To sustain the UV brightness, it is crucial that dusty outflows return to the galactic disk and re-activate star formation before massive stars with lifetimes of $\lesssim 100~\Myr$ cease producing UV radiation.

Indeed, the galaxy GS-z14-0 at $z = 14.32$, characterized by its bright UV luminosity ($M_{\rm UV} = -20.81~{\rm mag}$), and low dust attenuation ($A_V = 0.31$), may currently be in a "mini-quenching" phase of star formation, as suggested from the absence of strong emission lines \citep[][]{Carniani2024Natur, Ferrara2024A&A}.
If star formation is not reactivated within $\sim 100~\Myr$, the galaxy's UV luminosity is expected to decline below the JWST detection limit due to the depletion of its massive stars.

To assess whether dusty outflows can return to the galactic disk within this timescale, we estimate the free-fall timescale for dusty gas extending beyond the UV half-light radius by a factor of $\chi_{\rm d}$, i.e., $R_{\rm eff,d} = \chi_{\rm d}R_{\rm eff}$, which is given by:
\begin{eqnarray}
\tau_{\rm ff} = \sqrt{\frac{R_{\rm eff,d}^3}{G~M_{\rm g}}} \sim 20~\Myr~
\left ( \frac{M_{\rm g}}{10^{10}~\msun} \right )^{-1/2}
\left ( \frac{R_{\rm eff}}{500~\pc} \right )^{3/2}
\left ( \frac{\chi_{\rm d}}{5} \right )^{3/2} \ .
\label{eq:t_ff}        
\end{eqnarray}
For bright-end galaxies at $z \sim 12$, this estimated timescale is shorter than the lifetime of massive stars.
This suggests that even if star formation is temporarily halted by significant dust removal, it can resume through gas circulation before the UV luminosity declines noticeably.
This fact supports the plausibility of this dust displacement scenario.

\subsubsection{Large grain size}
The second possibility to reduce $\tau_{1500}$ is to decrease $\kappa_{1500}$ by assuming large dust grains with sizes of 0.1--1~$\micron$.
This scenario would be plausible in the early universe, where SNe are the dominant contributors to dust production, and small grains are preferentially destroyed by reverse shock in SN ejecta \citep[][]{Nozawa2007ApJ, Asano2013EP&S, Hirashita2019MNRAS}.
If the typical grain size, $a_{\rm g}$, is sufficiently larger than the reference UV wavelength ($\lambda = 0.15~\mu$m), the absorption cross-section can be approximated as \citep[][]{Ferrara2024arXiv}, 
\begin{eqnarray}
\kappa_{1500}^{\rm large} \sim \frac{\pi a_{\rm g}^2}{\frac{4\pi}{3} \delta_{\rm g} a_{\rm g}^3}
\sim 5085~{\rm cm^2 g^{-1}} \left ( \frac{a_{\rm g}}{0.5~\mu{\rm m}} \right )^{-1} \ ,
\label{eq:kappa_uv}        
\end{eqnarray}
where $\delta_{\rm g} = 2.95~{\rm g~cm^{-3}}$ is the material density of silicate grains, which are preferentially produced by SNe.
Thus, large grains with $a_{\rm g} = 0.5~\mu$m yield $\kappa_{1500}^{\rm large} \sim 0.04~\kappa_{1500}$, 
and the resulting $\tau_{\rm 1500}$ is of the same order as for the dust displacement scenario.
It is worth noting that a flat extinction curve, which arises due to such large grain sizes at wavelength shorter than $a_{\rm d}$, has been reported in $z > 6$ JWST-detected galaxies \citep[][]{Markov2024arXiv}.
Such flat attenuation curves would also be necessary to explain the blue excess slope observed in little red dots, which are promising candidates for $z > 5$ active galactic nuclei \citep[][]{Li2024arXiv}.
Thus, large dust grains would be a promising scenario for understanding various physical properties of galaxies in the early universe.

\subsubsection{Dust-to-star segregation}
Finally, we consider the third possibility: the spatial segregation of dust from young massive stars within the galactic disk.
Hydrodynamic simulations suggest that during the final stages of stellar cluster formation, stellar feedback clears internal gas and dust, leaving young stars within rarefied bubbles \citep[e.g.,][]{Hu2019MNRAS, Sugimura2024ApJ}.
This process allows a substantial fraction of UV light to escape from the star-forming regions effectively.
Indeed, recent galactic-scale simulations demonstrate that spatial segregation of dust from young stars reduces the overall dust attenuation experienced by galaxies \citep[e.g.,][]{Narayanan2018ApJ, Vijayan2024MNRAS}.
Unlike the dust displacement scenario discussed earlier, this mechanism allows substantial amounts of dust to remain in the galactic disk, potentially creating some highly obscured regions.
Consequently, the nature of dust attenuation predicted under this scenario can differ significantly from those predicted in the first two scenarios.
To explore this effect, we introduce a parameter $f_{\rm c}$, defined as the fraction of young massive stars embedded in optically thick gas. 
Assuming that UV photons from the remaining $(1-f_{\rm c})$ fraction of stars are completely unobscured, the observed UV luminosity can be expressed as,
\begin{eqnarray}
L_{\rm UV}^{\rm obs} = (1-f_{\rm c})L_{\rm UV}^{\rm int} + f_{\rm c} f_{\rm esc}(\tau_{1500}^{\prime}) L_{\rm UV}^{\rm int} \ ,
\label{eq:Luv_fc}        
\end{eqnarray}
where $\tau_{1500}^{\prime} \equiv \tau_{1500}/f_{\rm c}$ is the effective dust optical depth in the obscured regions, ensuring the total dust mass remains unchanged.

Figure~\ref{fig:LF_fc} shows UV LFs evaluated for different $f_{\rm c}$ values.
We find that the result with $f_{\rm c} \sim 0.1$ generally aligns with the observation at $z = 12$,
while $f_{\rm c} \sim 0.5$ is favorable for explaining the UV LF at $z = 7$.
Unlike models with reduced $\tau_{1500}$ (Figure~\ref{fig:LF}), 
the dust segregation model predicts a gradual decline in the bright-end of the UV LF,
as even massive galaxies with $\tau_{1500} > 100$ allow some UV photons to escape. 
Interestingly, the $f_{\rm c}$ values required to fit the observations decrease from $z = 12$ to $7$. 
This trend likely reflects the increasing mass of bright-end galaxies toward lower redshifts, which reduces the impact of stellar feedback on the ISM structure due to their deeper gravitational potential wells.

Previous studies have however argued that dust segregation enhances FIR fluxes, potentially conflicting with the observed non-detection of dust continuum in $z > 10$ galaxies.
\citet{Ferrara2022MNRAS} demonstrated that in the absence of dust segregation, FIR fluxes have an upper limit for given UV fluxes and dust extinction. 
A significant excess of FIR fluxes above this theoretical upper limit would indicate substantial dust-to-star segregation in galaxies.
Following this argument, \citet{Ziparo2023MNRAS} suggested that $z > 10$ galaxies are unlikely to have strong dust segregation, as their intrinsic FIR fluxes inferred from the non-detection have to be far below the theoretical upper limits.
Furthermore, it is highly ambiguous whether significant segregation ($f_{\rm c} \sim 0.1$) is achievable in the bright-end galaxies at $z > 10$, particularly given their very compact nature.
Their high surface densities ($\sgmg \sim 10^4~\msun \pc^{-2}$) could hinder effective bubble formation by stellar feedback \citep[e.g.,][]{Grudic2018MNRAS, Dekel2023MNRAS}.
Thus, whether dust segregation can self-consistently explain the observed properties of $z > 10$ galaxies remains an open question.
Addressing this issue requires high-resolution hydrodynamic simulations capable of resolving sub-pc structures in dusty ISM while incorporating both UV and FIR radiation transfer in galaxies.

\subsection{Weak-feedback scenario}\label{sec:fbf}

Our fiducial model that assumes the KS law with $\fboost = 1$ and a high mass loading factor of $\fout = 5$ likely provides a conservative prediction on SFEs of $z > 5$ galaxies.
To explore a more optimistic scenario, we consider a weak-feedback (WFB) model, where the SFR per unit gas mass increases and the mass loading factor decreases in environments of higher gas densities.
This model is motivated by radiation hydrodynamics simulations of \citet{Fukushima2021MNRAS} showing a sharp rise in the SFR per cloud when the cloud surface density exceeds a critical threshold, $\Sigma_{\rm cr}$.
To implement this effect in our model, we express the boost factor $\mathcal{F}_{\rm b}$ in Eq.~(\ref{eq:sgmsf}) as follows:
\begin{eqnarray}
\mathcal{F}_{\rm b} = 1 + \frac{\mathcal{F}_{\rm b,0}-1}{1+{\rm exp}(-f_{\rm cr})} \ , \ \ f_{\rm cr} = 5 \frac{\sgmg-\Sigma_{\rm cr}}{\Sigma_{\rm cr}} \ .
\label{eq:fboost}
\end{eqnarray}
This expression refers to Eqs.~(16) and (17) of \citet{Fukushima2021MNRAS} and leads to $\mathcal{F}_{\rm b} \sim 1$ at $\sgmg \ll \Sigma_{\rm cr}$
and $\mathcal{F}_{\rm b} \sim \mathcal{F}_{\rm b, 0}$ at $\sgmg \gg \Sigma_{\rm cr}$.
We set $\mathcal{F}_{\rm b,0} = 10$ according to spatially-resolved SFRs measured in $z > 5$ galaxies \citep[][]{Vallini2024MNRAS}.
We here assume $\Sigma_{\rm cr} = 10^3~\msun \pc^{-2}$, above which the gravitational potential of the clouds is considered to be deep enough to confine gas accelerated by supernova momentum deposition \citep[e.g.,][]{Grudic2018MNRAS, Grudic2020MNRAS, Byrne2023MNRAS}.
For consistency with the increased SFRs, we depress the mass loading factor, scaling it by a factor of $( 1 + \sgmg / \Sigma_{\rm cr} )^{-1}$.

Our WFB model is modest compared to the feedback-free model proposed by \citet{Dekel2023MNRAS}, which assumes the complete absence of outflows in star-forming regions with high gas densities of $\gtrsim 10^{3}~{\rm cm^{-3}}$. 
They suppose that, in such dense environments, the free-fall time of gas clouds is significantly shorter than the lifetimes of massive stars, allowing a substantial fraction of gas to be converted into stars before supernova-driven feedback operates effectively.
Thus, if our WFB model already overpredicts key quantities, such as SFRs, stellar masses, gas metallicities, and dust masses, relative to observational constraints, the feedback-free model would inevitably lead to even larger discrepancies.

Figure~\ref{fig:ms_sfr_z_fbf} summarizes several key results from the WFB model, showing the SFR, metallicity, and dust mass as functions of stellar mass.
Compared to the fiducial model, the WFB model predicts higher SFRs, particularly in massive halos with $M_{\rm h}(z = 5) \gtrsim 10^{12}~\msun$, where disk surface densities are sufficiently higher than $\Sigma_{\rm cr} = 10^3~\msun \pc^{-2}$.
For these massive halos, metallicities are elevated by a few 10~$\%$, compared to the fiducial model, owing to the suppression of metal evacuation.
This enhanced metallicity accelerates dust mass growth via metal accretion,
leading to dust masses a few times higher than those in the fiducial model.

\begin{figure}
\centering
\includegraphics[width=0.9\columnwidth]{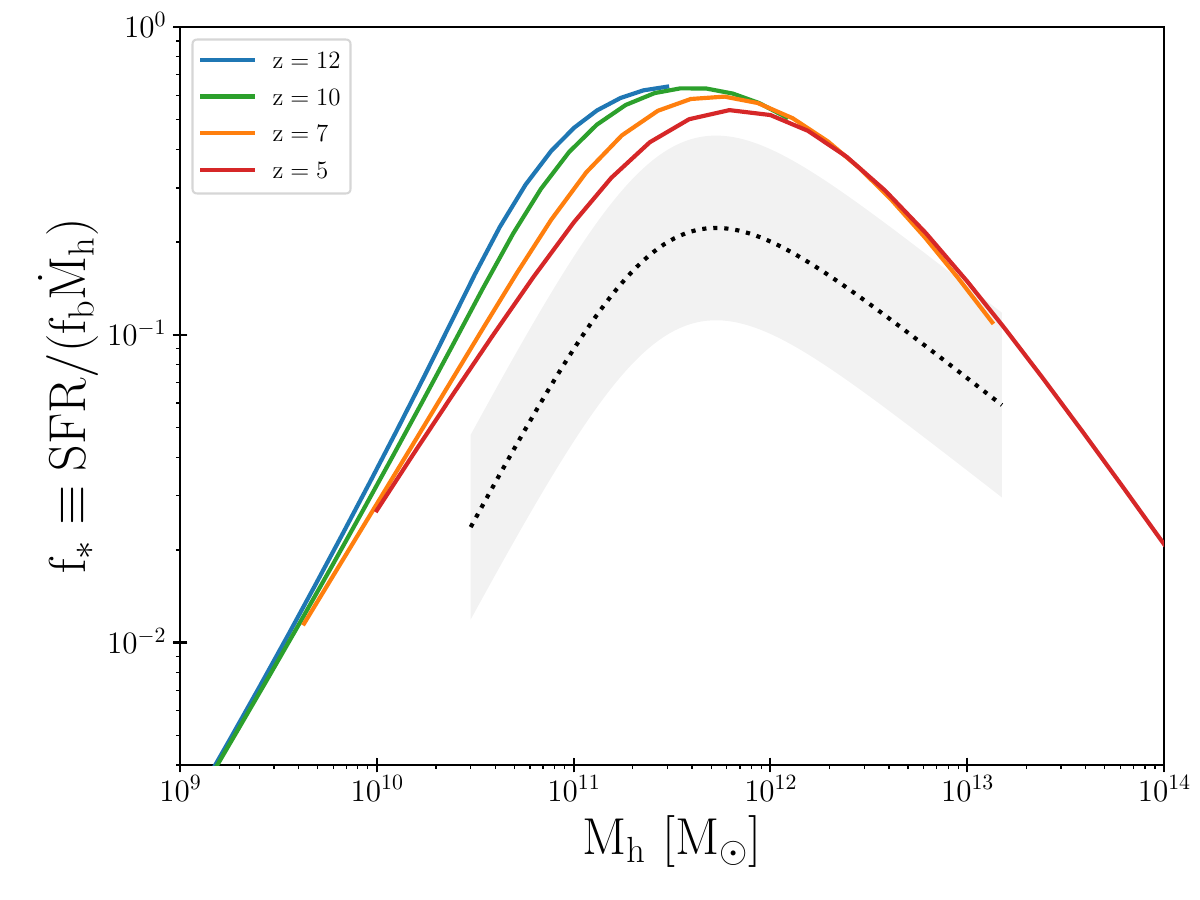}
\caption{
The same as Figure~\ref{fig:SFE}, but showing the result of the weak feedback model.
}
\label{fig:SFE_fbf}
\end{figure}

The impact of weak feedback on SFEs is examined in Figure~\ref{fig:SFE_fbf}.
The $f_\ast$--$M_{\rm h}$ relation in the WFB model is systematically shifted upward compared to observations.
In particular, the peak SFE reaches $f_\ast \sim 0.6$, which is approximately three times greater than the observed value.
Moreover, the redshift dependence of the $f_\ast$--$M_{\rm h}$ relation is more pronounced than in the fiducial model.
This is because, in the WFB model, the mass loading factor evolves over time in response to changing surface gas densities.
For comparison, we note that the feedback-free model predicts $f_\ast \gtrsim 0.5$ 
for all halos with $M_{\rm h} \gtrsim 10^{10}~\msun$ at $z \sim 12$ \citep[][]{Li2024A&A}, 
although our model provides such high SFEs only for halos with $M_{\rm h} \sim 10^{11-12}~\msun$.

\begin{figure}
\centering
\includegraphics[width=0.9\columnwidth]{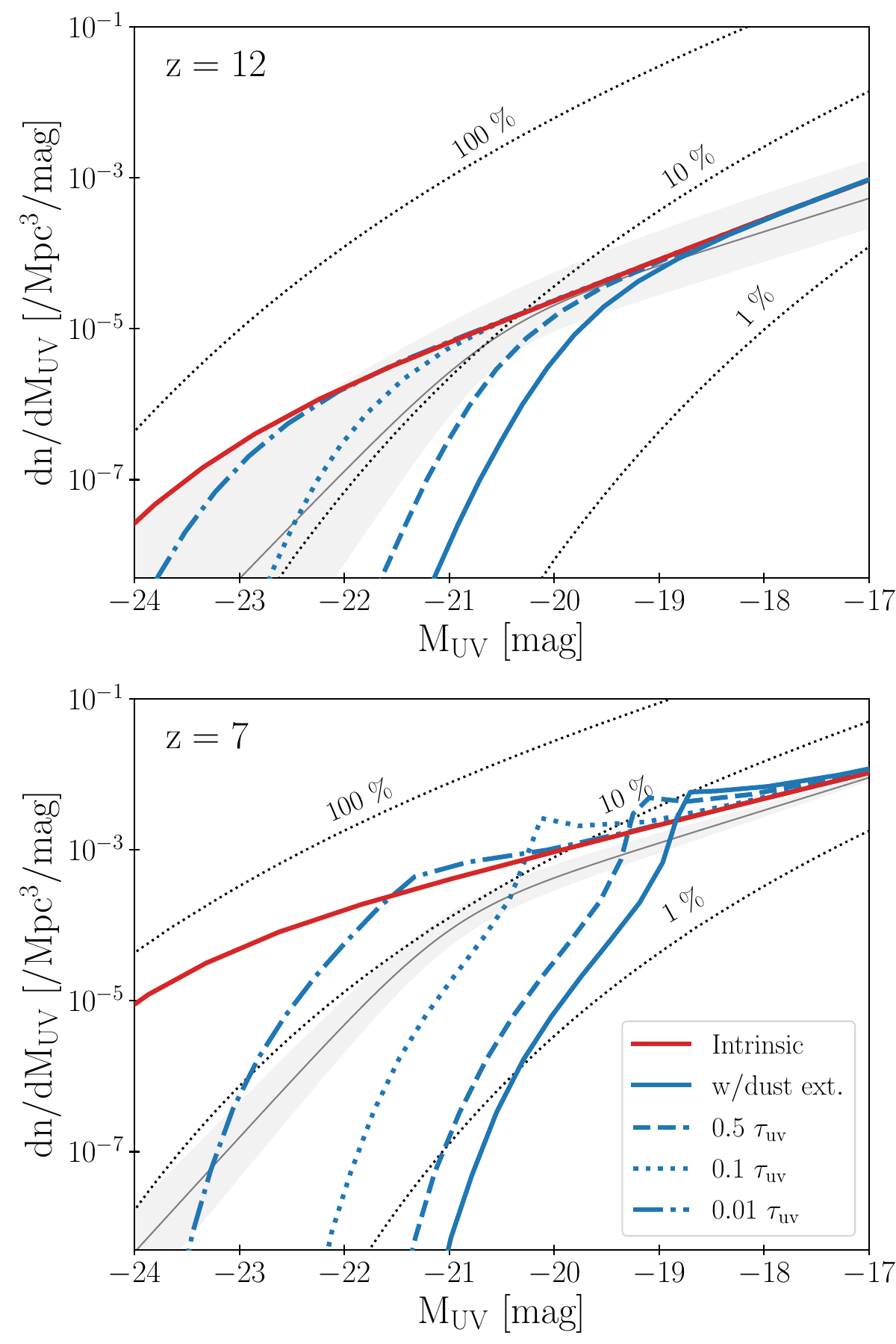}
\caption{
The same as Figure~\ref{fig:LF}, but showing the UV LFs predicted by the weak feedback model.
}
\label{fig:LF_fbf}
\end{figure}

As shown in Figure~\ref{fig:LF_fbf}, the elevated SFEs in the WFB model extend the bright end of the intrinsic UV beyond those predicted by the fiducial model.
However, the dust-attenuated UV LFs remain similar to those in the fiducial model, indicating that the WFB model leads to stronger dust attenuation due to efficient dust growth facilitated by suppressed metal evacuation.
Therefore, even under the WFB scenario, mitigating dust attenuation remains crucial to describe the galaxy evolution at such high redshifts.

\begin{figure}
\centering
\includegraphics[width=0.9\columnwidth]{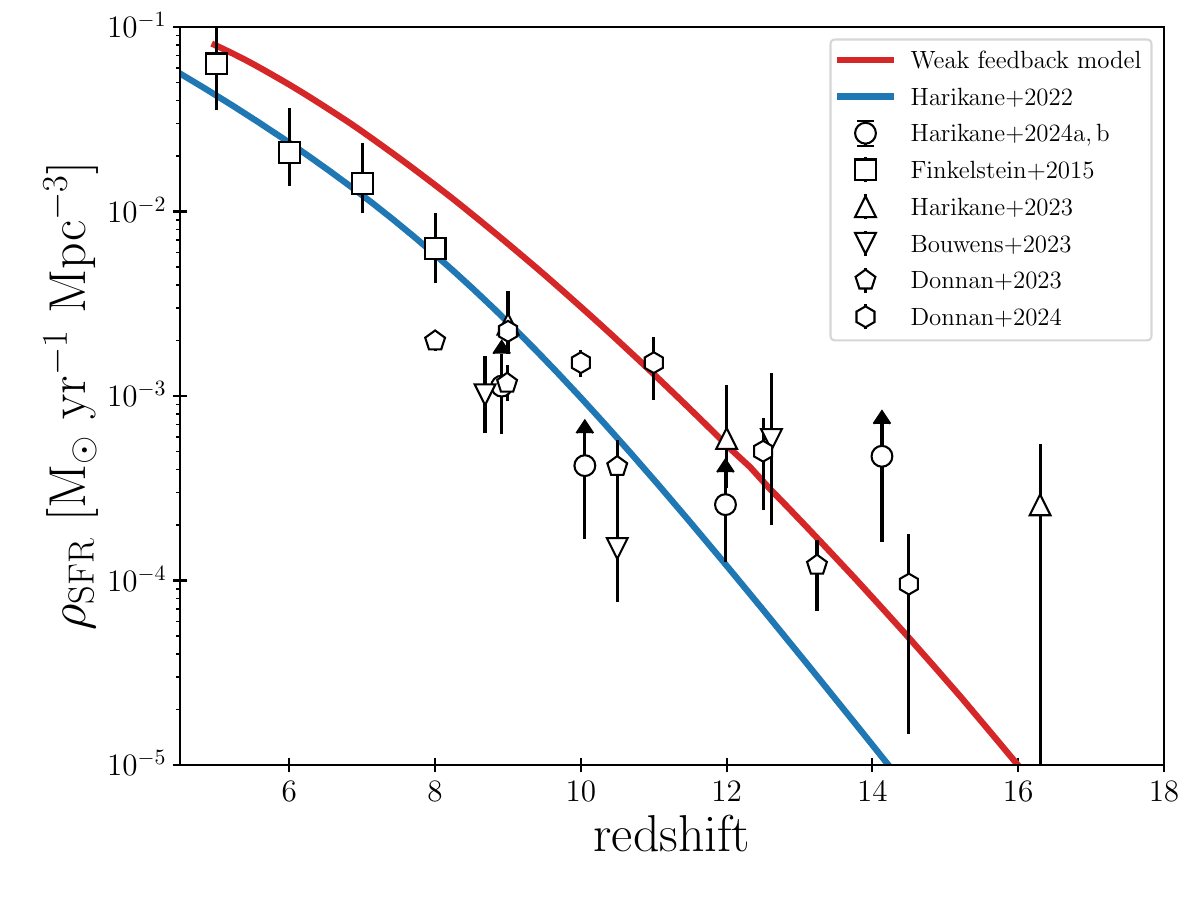}
\caption{
The same as Figure~\ref{fig:CSFR}, but showing the result of the weak feedback model.
}
\label{fig:CSFR_fbf}
\end{figure}

Finally, we examine the CSFRD predicted by the WFB model in Figure~\ref{fig:CSFR_fbf}.
The WFB model generally reproduces the observed CSFRDs at $z \sim 12$-14.
However, at $z < 10$, it significantly overestimates CSFRDs, 
suggesting that the WFB phase of galaxies ceases around $z \gtrsim 12$, after which SFEs become more strongly regulated by galactic outflows.
Understanding the mechanisms that terminate the WFB phase is an important subject for future research.

%%%%%%%%%% fig:SFE_fbf %%%%%%%%%%%%%%%%
%%%%%%%%%%%%%%%%%%%%%%%%%%%%%%%%%%

%%%%%%%%%% fig:LF_fbf %%%%%%%%%%%%%%%%
%%%%%%%%%%%%%%%%%%%%%%%%%%%%%%%%%%

%%%%%%%%%% fig:CSFR_fbf %%%%%%%%%%%%%%%%
%%%%%%%%%%%%%%%%%%%%%%%%%%%%%%%%%%

% %%%%%%%%%% fig:FIR_fbf %%%%%%%%%%%%%%%%
% \begin{figure}
% \centering
% \includegraphics[width=0.9\columnwidth]{figure/FIR_Mhalo_fbf.pdf}
% \caption{
% }
% \label{fig:FIR_fbf}
% \end{figure}
% %%%%%%%%%%%%%%%%%%%%%%%%%%%%%%%%%%

%%%%%%%%%%%%%%%%%%%%%%%%%%%%%%%%%%%%%%%%%%%%%%%%%%

\section{Summary}\label{sec:summary}

In this paper, we have presented a new galaxy evolution model specialized for the early universe from $z = 20$ to $5$.
Our model self-consistently calculates the radially-resolved mass evolution of gas, stars, heavy elements, and dust for galaxies with different dark-matter halo masses.
In these calculations, we incorporate variations in the spatial extent of gas inflows ($\fspin$), the mass loading factor of gas outflows ($\eta$), and the timescale for dust growth ($\tau_{\rm acc}$).
With reasonable parameter choices, our model successfully reproduces various observed properties of distant star-forming galaxies identified by JWST at $z > 5$. 
The main findings of this study are summarized as follows:
\begin{itemize}
\setlength{\itemsep}{0.2cm}

\item The decreasing trend of UV half-light radii of galaxies for higher-redshifts is naturally explained by assuming that the spatial extent of gas inflows scales with the virial radius ($h_{\rm R} = \fspin \lambda_{\rm s} r_{\rm vir}$).
The observed dispersion in galaxy sizes likely results from the diversity in the $\fspin$ values, which are inferred to vary in the range of $\fspin = 0.1$--$2$.
This variation could result from angular momentum redistribution in baryons, driven by disk instabilities and/or galaxy mergers.

\item Substantial outflows characterized by a mass loading factor of $\fout \sim 5$ are necessary to explain the observed metallicity for $z > 5$ galaxies, approximately a few times lower than the local scale relation.
Similar to $\fspin$, the $\fout$ values likely vary in the range $\fout = 1$--$100$ to reproduce the diversity in the observed metallicities.

\item A moderately short timescale for dust growth given by $\tau_{\rm acc} = 5~\Myr~(Z/\zsun)^{-1}$ is favorable for the formation of dust-rich galaxies observed at $z \sim 7$. 
This timescale allows rapid dust mass growth at $z < 8$, which is facilitated by galaxy chemical enrichment, while remaining consistent with the observational upper limit in dust mass for $z > 10$ galaxies, such as GN-z11, GHZ2, and GS-z14-0.

\item We have investigated star formation efficiencies (SFEs), defined as $f_\ast \equiv \sfr / (f_{\rm b} \dot{M}_h)$, and found that the $f_\ast$--$M_{\rm h}$ relation is nearly redshift-independent.
This relation peaks at $f_\ast \sim 0.2$ for halos with $M_{\rm h} \sim 10^{12}~\msun$, following scaling relations of $f_\ast \propto M_{\rm h}$ for lower-mass halos and $f_\ast \propto M_{\rm h}^{-1}$ for higher-mass halos.
The value of $f_\ast$ is primarily regulated by the global balance between gas supply and consumption within galaxies, leading to a negative correlation between $f_\ast$ and $\fout$.
Our predicted $f_\ast$--$M_{\rm h}$ relation closely matches observations at $z \sim $2--7,
suggesting a non-evolving mass loading factor of $\fout \sim 5$ across a broad redshift range.

\item The intrinsic UV luminosity function predicted by our model at $z = 12$ generally aligns with the observed one.
This agreement arises from the high star formation efficiency mentioned above.
However, our model galaxies are considerably dust-obscured due to high optical depths ($\tau_{1500} \gtrsim 10$), causing the dust-attenuated UV LF to fall far below the observed one.
This trend is also confirmed at $z = 7$.
These results highlight the need for potential mechanisms that effectively mitigate such significant dust attenuation over a wide redshift range.

\item To address the above point, we discussed three possible scenarios: (a) dust displacement, (b) large grain size, and (c) dust-to-star spatial segregation.
The quantitative analysis suggests that all three scenarios can reconcile the UV LFs predicted by our model with the observed ones without significant changes in our model description.
Thus, we emphasize the crucial importance of further exploring these dust processes in future theoretical studies to fully understand the observed properties of galaxies at $z > 5$.

\item We have also explored a weak feedback (WFB) model, in which SFRs are enhanced, and outflow rates are suppressed in dense environments with $\sgmg > 10^3~\msun~\pc^{-2}$.
This model predicts a peak SFE of $f_\ast \sim 0.6$, approximately three times higher than in the fiducial model, leading to higher intrinsic UV luminosities.
However, as in the fiducial case, the WFB model fails to reproduce the observed UV LFs at $z > 5$, since reduced dust evacuation results in stronger dust attenuation.
Thus, rather than simply increasing SFEs, mitigating dust attenuation is more crucial for resolving the discrepancy between our model predictions and the observed UV LFs in the early universe.

\end{itemize}

We emphasize that our model represents the statistically averaged nature of galaxy evolution.
In reality, galaxies undergo discontinuous mergers, leading to intermittent star formation and mass evacuation through outflows, thereby establishing the diverse characteristics observed in high-$z$ galaxies.
Stochastic feedback events also play an essential role in triggering dust displacement and dust segregation processes.
Therefore, validating our model prediction with cosmological galaxy formation models and simulations that incorporate these dust processes is a crucial task for achieving a more comprehensive understanding of galaxy evolution.

%%%%%%%%%%%%%%%%%%%%%%%%%%%%%%%%%%%%%%%%%%%%%%%%%%

\section*{Acknowledgements}

We thank M.~Ouchi, H.~Hirashita, L.~Romano, and M.~Kohandel for fruitful discussions. 
D.\ T.\ was supported in part by the JSPS Grant-in-Aid for Scientific Research (21K20378). 
H.\ Y.\ was supported by the MEXT/JSPS KAKENHI (21H04489) and the JST FOREST Program (JP-MJFR202Z).
This work is supported by the  JSPS International Leading Research (ILR) project, JP22K21349. 
K. N. is supported by the JSPS KAKENHI grant 24H00002, 24H00241.  
K. N. acknowledges the support from the Kavli IPMU, the World Premier Research Center Initiative (WPI), UTIAS, the University of Tokyo.
AF acknowledges support from the ERC Advanced Grant INTERSTELLAR H2020/740120. 
This research was supported (AF) in part by grant NSF PHY-2309135 to the Kavli Institute for Theoretical Physics (KITP). 

% K.\ I.\ acknowledges support from the National Natural Science Foundation of China (12073003, 12003003, 11721303, 11991052, and 11950410493) and the China Manned Space Project (CMS-CSST-2021-A04 and CMS-CSST-2021-A06). 
% R.\ K.\ acknowledges financial support via the Heisenberg Research Grant funded by the Deutsche Forschungsgemeinschaft (DFG, German Research Foundation) under grant no.~KU 2849/9, project no.~445783058.
% The numerical simulations were performed with the Cray XC50 at the Center for Computational Astrophysics (CfCA) of the National Astronomical Observatory of Japan.
% This work is supported in part by XXXX, YYYY, and ZZZZ.
% JSPS KAKENHI Grant Numbers 21K20378, 
% the National Natural Science Foundation of China (12073003, 12003003, 11721303, 11991052, 11950410493, and 1215041030), 
% the China Manned Space Project Nos. CMS-CSST-2021-A04 and CMS-CSST-2021-A06, 
% NSF grant AST-2006176, 
% the Heisenberg Research Grant funded 
% %by the German Research Foundation 
% No.~KU 2849/9, 
% and the JSPS Invitational Fellowship for Research in Japan ID S20156.

%%%%%%%%%%%%%%%%%%%%%%%%%%%%%%%%%%%%%%%%%%%%%%%%%%

\section*{Data availability}

The data underlying this article will be shared on reasonable request to the corresponding author.

%%%%%%%%%%%%%%%%%%%%%%%%%%%%%%%%%%%%%%%%%%%%%%%%%%



\bibliographystyle{mnras}
\bibliography{refs.bib} % if your bibtex file is called example.bib

%\begin{thebibliography}{99}
%\end{thebibliography}

\appendix

% \section{Effects of varying initial rotational velocity}\label{sec:vr}

\section{$z \gtrsim 10$ galaxies identified by JWST}\label{sec:obs}

%%%%%%%%%%%%%%%%%%%%%%%%%%%%%%%%%%
\renewcommand{\arraystretch}{1.5}
\begin{table*}
\centering
\begin{tabular}{cccccccc}
\hline \hline
Name & $z$ & $\rm M_{UV}$ & $\rm log(\mstar)$ & $\sfr$ & $\oh$ & $R_{\rm eff}$ & Reference \\
 & & $\rm [mag]$ & $\rm [M_\odot]$ & $\rm [M_\odot yr^{-1}]$ & & $\rm [pc]$ & \\
\hline
GS-z14-0    & 14.32 & $-20.81$ & $8.6^{+0.7}_{-0.2}$    & $19\pm6$               & $7.94\pm0.03$ & $260\pm20$ & 1,2 \\
GS-z14-1    & 13.9  & $-19.0$  & $8.0^{+0.4}_{-0.3}$    & $2^{+0.7}_{-0.4}$      & $-$ & $< 160$ & 1 \\
GS-z13      & 13.2  & $-18.92$ & $7.7^{+0.4}_{-0.2}$    & $1.4^{+0.6}_{-0.2}$    & $-$ & $59$ & 3 \\
UNCOVER-z13 & 13.08 & $-19.4$  & $8.13^{+0.11}_{-0.15}$ & $1.28^{+0.27}_{-0.18}$ & $-$ & $309^{+110}_{-74}$ & 4 \\
GS-z13-1-LA & 13.05 & $-18.7$  & $7.74^{+0.15}_{-0.12}$ & $0.16^{+0.48}_{-0.15}$ & $-$ & $62^{+11}_{-6}$ & 5 \\
GS-z12      & 12.48 & $-18.8$  & $7.68\pm0.19$          & $1.62^{+0.28}_{-0.24}$ & $7.59\pm0.25$ & $146\pm11$ & 6 \\
UNCOVER-z12 & 12.39 & $-19.2$  & $8.35^{+0.18}_{-0.14}$ & $2.15^{+0.81}_{-0.46}$ & $-$ & $426^{+40}_{-42}$ & 4 \\
GHZ2        & 12.34 & $-20.53$ & $8.27^{+0.23}_{-0.18}$ & $8.7^{+1.4}_{-1.9}$    & $7.26^{+0.27}_{-0.24}$ & $34\pm9$ & 7,8 \\
Maisie      & 11.42 & $-20.1$  & $8.6\pm0.3$            & $2\pm1$                & $8$ & $280^{+80}_{-62}$ & 9 \\
GS-z11      & 11.12 & $-19.32$ & $8.3^{+0.1}_{-0.1}$    & $1.45\pm0.1$           & $-$ & $119$ & 3 \\
CEERS2\_588 & 11.04 & $-20.4$  & $9.0^{+0.5}_{-0.2}$    & $12.7^{+9.7}_{-4.9}$   & $-$ & $476$ & 10 \\
GN-z11      & 10.60 & $-21.5$  & $8.73\pm0.06$          & $18.8^{+0.81}_{-0.69}$ & $7.84^{+0.06}_{-0.05}$ & $64\pm20$ & 11 \\
UHZ-1       & 10.07 & $-19.85$ & $8.14^{+0.09}_{-0.11}$ & $1.25^{+0.18}_{-0.12}$ & $-$ & $592\pm85$ & 12 \\
CEERS\_35590 & 10.01 & $-20.1$ & $9.1\pm0.1$            & $9\pm2$                & $-$ & $420\pm20$ & 13 \\
CEERS\_99715 &  9.97 & $-20.5$ & $9.5\pm0.1$            & $6^{+4}_{-2}$          & $-$ & $580\pm20$ & 13 \\
\hline \hline
\end{tabular}
\caption{Observational data for star-forming galaxies at $z \gtrsim 10$. 
1. \citet{Carniani2024arXiv}, 
2. \citet{Carniani2024Natur}, 
3. \citet{Hainline2024arXiv},
4. \citet{Wang2023ApJ},
5. \citet{Witstok2024arXiv},
6. \citet{D'Eugenio2024},
7. \citet{Castellano2024ApJ},
8. \citet{Mitsuhashi2025arXiv},
9. \citet{Arrabal_Haro2023bNatur},
10. \citet{Harikane2024aApJ},
11. \citet{Bunker2023AA},
12. \citet{Goulding2023ApJ},
13. \citet{Arrabal_Haro2023aApJ}
}
\label{table:sample}
\end{table*}
\renewcommand{\arraystretch}{1}
%%%%%%%%%%%%%%%%%%%%%%%%%%%%%%%%%%

%%%%%%%%%%%%%%%%%%%%%%%%%%%%%%%%%%%%%%%%%%%%%%%%%%

% Don't change these lines
\bsp	% typesetting comment
\label{lastpage}
\end{document}